\documentclass[fleqn,usenatbib]{mnras}

\usepackage{newtxtext,newtxmath}
\usepackage[dvipsnames]{xcolor}

\usepackage[T1]{fontenc}

\DeclareRobustCommand{\VAN}[3]{#2}
\let\VANthebibliography\thebibliography
\def\thebibliography{\DeclareRobustCommand{\VAN}[3]{##3}\VANthebibliography}

\usepackage{graphicx}	%
\usepackage{amsmath}	%
\usepackage{booktabs}
\usepackage[normalem]{ulem} %
\usepackage{subcaption}
\usepackage{float}
\usepackage{grffile}

\usepackage{array}
\newcolumntype{x}[1]{>{\centering\let\newline\\\arraybackslash\hspace{0pt}}p{#1}}

\graphicspath{{figures/}}

\newcommand{\Msun}{\text{M}$_{\sun}$}
\newcommand{\Hi}{H\textsc{i}}

\title[A tale of two tails]{A tale of two tails: insights from simulations into the formation of the peculiar dwarf galaxy NGC 1427A}

\author[M. Mastropietro et al.]{
M. Mastropietro,$^{1}$\thanks{E-mail: michele.mastropietro@ugent.be}
S. De Rijcke,$^{1}$
R. F. Peletier.$^{2}$
\\
$^{1}$Department Physics and Astronomy, Ghent University, Krijgslaan 281, S9, B-9000 Gent, Belgium\\
$^{2}$Kapteyn Astronomical Institute, University of Groningen, PO Box 800, NL-9700 AV Groningen, the Netherlands\\
}

\date{Accepted XXX. Received YYY; in original form ZZZ}

\pubyear{2020}

\begin{document}
\label{firstpage}
\pagerange{\pageref{firstpage}--\pageref{lastpage}}
\maketitle

\begin{abstract}
We present a scenario for the formation and the morphology of the arrow-shaped dwarf irregular galaxy NGC~1427A in the Fornax Cluster.
This galaxy shows intriguing stellar and gaseous tails pointing in different directions for which alternative but not conclusive formation scenarios have been proposed in the literature.
We performed $N$-body/SPH simulations of dwarf galaxies falling into a model of the Fornax cluster, exhibiting a jellyfish-like appearance while undergoing ram-pressure stripping.
We noted that some of our models show interesting tail morphologies similar to that of NGC~1427A.
In this way, the peculiar NGC~1427A structure can be studied using models whose stellar and neutral gas photometry and kinematics are in good agreement with the observed ones, without the need of invoking an interaction with a nearby galaxy.
Thanks to the tails, we can identify the requirements for a galaxy to expose such a structure and assess the possible position and velocity of the galaxy in the cluster.
This puts constraints on the orbit of the galaxy, its position in the cluster and the time since its pericentre passage.
From the statistics of identified snapshots following our modelling, we found that the most likely position of the galaxy is around $200$~kpc in front of the cluster centre, travelling towards the cluster with a velocity angle with respect to the line-of-sight direction of around $50$~deg.
This analysis can be useful in future observations of similar galaxies in clusters to characterise their position and velocity in the cluster and their formation.
\end{abstract}


\section{Introduction} \label{sec:intro}

\begin{figure*}
\centering
\includegraphics[width=\textwidth]{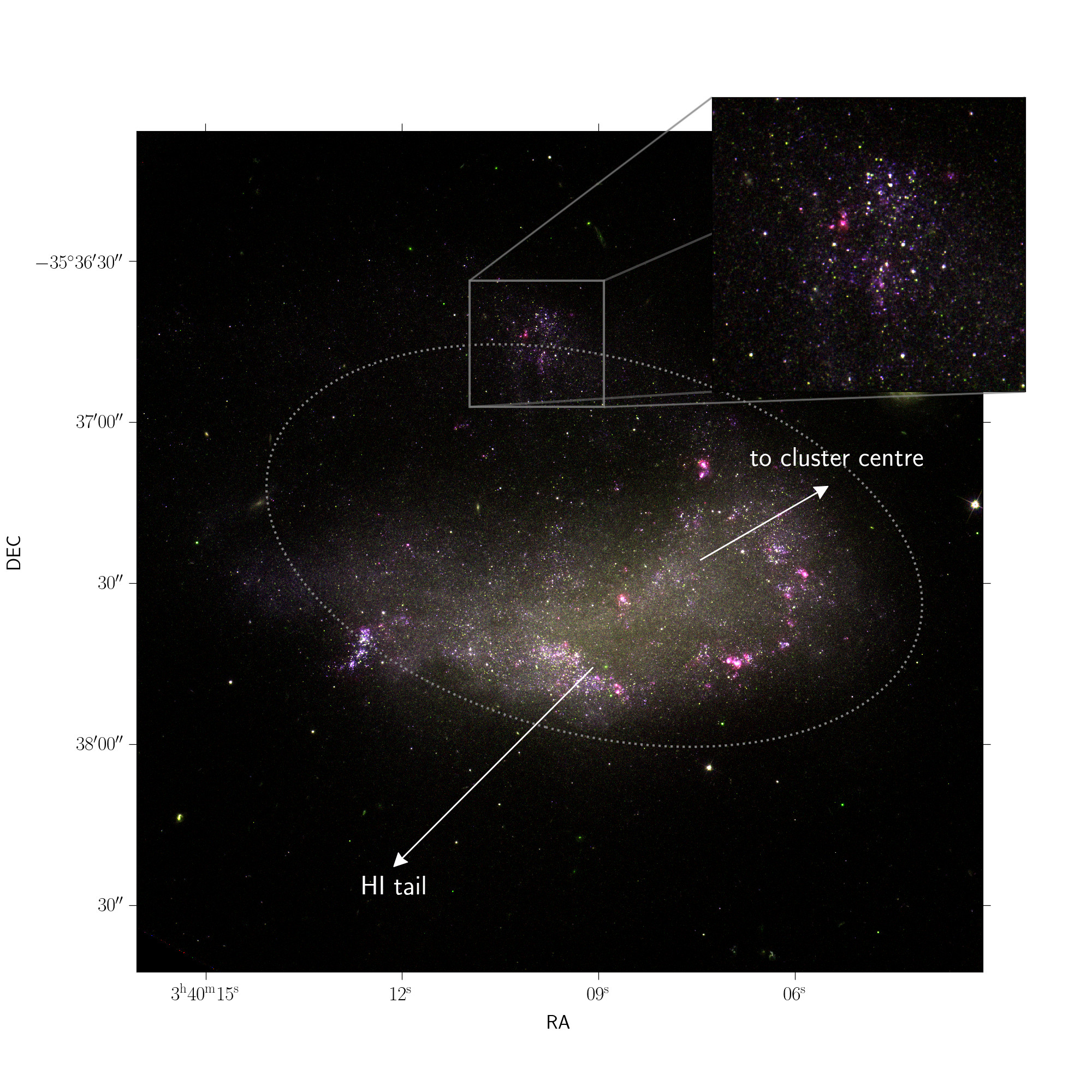}
\caption{False-colour image of NGC 1427A, based on HST Advanced Camera for Surveys (ACS) archival data (Proposal ID: 9689; PI: M. Gregg).
The following colour bands were used: blue=475W; green=775W; red=625W+660N (stellar emission was subtracted from the 660N image using a scaled 775W image).
This colour scheme makes the H$\alpha$ emission stand out in red. An asinh stretch was applied to bring out also the faint details.
The inset zooms in on the Northern Clump, showing it to be composed of two loose stellar clusters with embedded H{\sc ii} emission.
The Northern Clump also appears to be connected to the north-west rim of NGC 1427A's main body via a tenuous stream of stars.
The directions of the \Hi{} tail and towards the Fornax Cluster centre are indicated with arrows.
The dotted ellipse is the same as in Figures 1 and 2 of \citet{Lee-Waddell2018} and indicates the shape and direction of the faint outskirts of the galaxy, which are quite distinct from the system's inner, brighter parts.
}
\label{fig:NGC1427A}
\end{figure*}

The evolution of galaxies in dense environments has been shown to be markedly different from that of more isolated galaxies, with mass being a prominent factor in determining how profoundly environmental influences affect a galaxy \citep{Boselli2006, Grossi2018a}.
The well-known morphology-density relation, according to which early-type galaxies are mostly found in high-density environments \citep{Dressler1980, Dressler1997}, is especially pronounced for low-mass systems, such as dwarf galaxies \citep{McConnachie2012}.
Indeed, while actively star-forming late-type dwarf galaxies are found almost exclusively in low-density environments, truly isolated quiescent early-type dwarf galaxies, on the contrary, are exceedingly rare \citep{Binggeli1990, Karachentseva2010, Geha2012}.

An effective way of shutting down the star formation in a galaxy is to rob it of the raw material for building stars:~gas.
When a galaxy enters on an orbit in a galaxy cluster or group, it is subjected to the tidal forces of the cluster potential and of its galaxies.
Its interstellar medium experiences the ram pressure \citep{GunnGott1972}, basically a supersonic "headwind", exerted by the intracluster medium.
Ram pressure is a well known phenomenon and many studies have been devoted to simulating its effects on galaxies \citep[e.g.:][]{Mori2000, Mayer2006, Roediger2008, Roediger2015, Steinhauser2016, Yun2018, Steyrleithner2020}.
If the ram pressure is sufficiently vigorous, the galaxy's diffuse interstellar medium can be pushed out of its gravitational well, forming a tail of escaping \Hi{} gas in the galaxy's wake.
The much more clumpy molecular gas is not as easily removed by the ram pressure and remains behind while being consumed by star formation \citep{Abramson2014, Lee2017, Wang2020}.
Inside the tail, gas can cool and form knotty condensations, leading to a complex stellar system, with a head consisting of the galaxy's stellar body (and what remains of its gas) and a tail of twisting swirls of stripped gas, beaded with knots of star formation.

The term \emph{jellyfish galaxies} \citep{Ebeling2013} neatly fits this description and hence they are prime candidates for interpreting the transformation processes acting on galaxies in cluster and group environments.
Jellyfish galaxies exhibit tentacles of material that appear to be stripped from the galaxy body \citep{Poggianti2017a, Poggianti2019b, Ramatsoku2020}.
The term applies to galaxies with star formation activity within the stripped gas tails.
Signatures of the newly born stars within those gaseous tails are easily found in UV or blue images \citep{Cortese2007,Smith2010a,Poggianti2017a}. In particular, as will be explored more in detail in another study on the same simulations presented in this paper (Mastropietro et al., in prep.), star formation flickers on and off inside the gaseous tails of simulated ram-pressure stripped dwarf galaxies as gas clumps condense and disperse again. This gives the label "jellyfish galaxy" a transient and possibly recurrent quality.

The distorted optical appearance and gaseous tails and overall jellyfish-like morphology -- barring currently active star formation in the gas tails --  of NGC 1427A (Figure \ref{fig:NGC1427A}) seems to be straightforwardly and satisfactorily explained by ram-pressure stripping in conjunction with cluster tidal forces.
To correctly interpret the available data, it is of prime importance to be able to reliably identify the dominant transformation process for this galaxy.
With this goal in mind, we compare recent \Hi{} and optical data of NGC 1427A with a suite of dwarf galaxy simulations, set in a Fornax Cluster environment.

In the next section, we give a short overview of the observed properties of NGC 1427A and focus on those that are considered most relevant for elucidating its origin and evolution. In Section \ref{sec:simulations}, we present the numerical details of our simulations. The methodology behind the comparison of these simulations with the observations is discussed in Section \ref{sec:results}. We conclude with a discussion of the main results in Section \ref{sec:discussion}.

\section{Observed properties of NGC 1427A} \label{sec:observations}
\begin{figure*}
\begin{subfigure}[b]{0.33\textwidth}
  \centering
  \includegraphics[width=\textwidth]{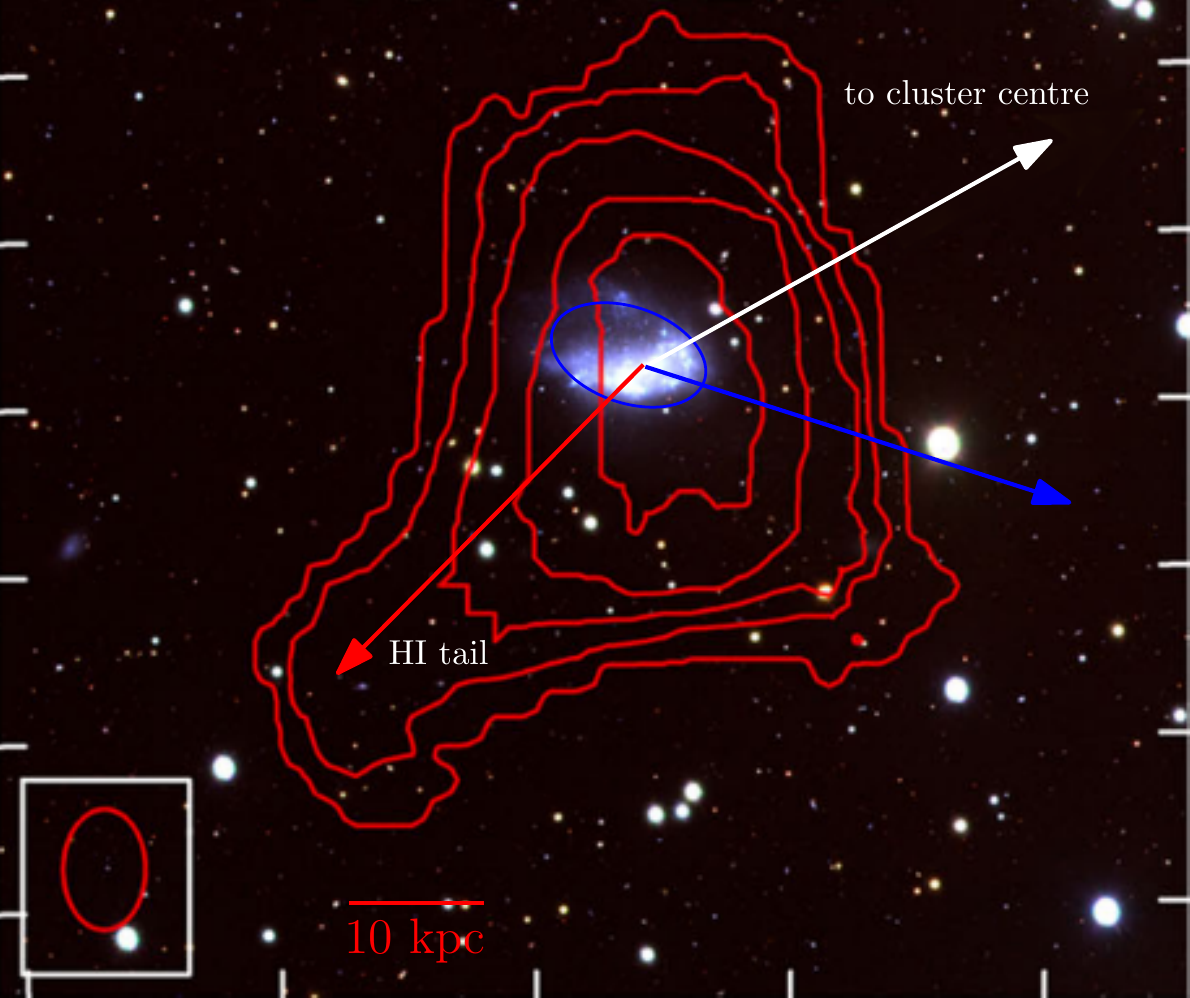}\\[3ex]
  \caption{}
  \label{fig:hi_contours}
\end{subfigure}
\begin{subfigure}[b]{0.33\textwidth}
  \centering
  \includegraphics[width=0.9\textwidth]{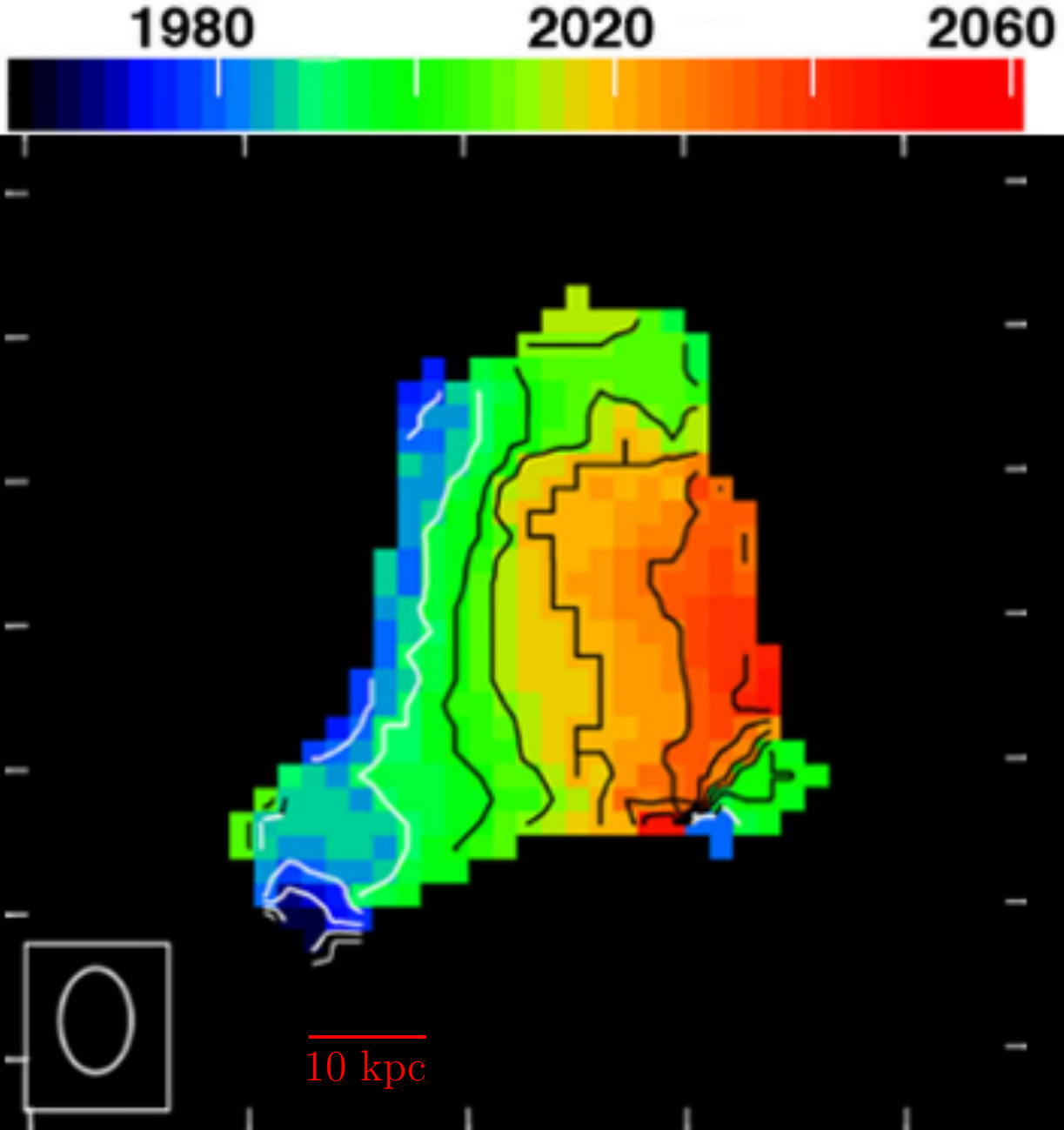}\\[3ex]
  \caption{}
  \label{fig:hi_kin}
\end{subfigure}
\begin{subfigure}[b]{0.33\textwidth}
  \centering
  \includegraphics[width=\textwidth]{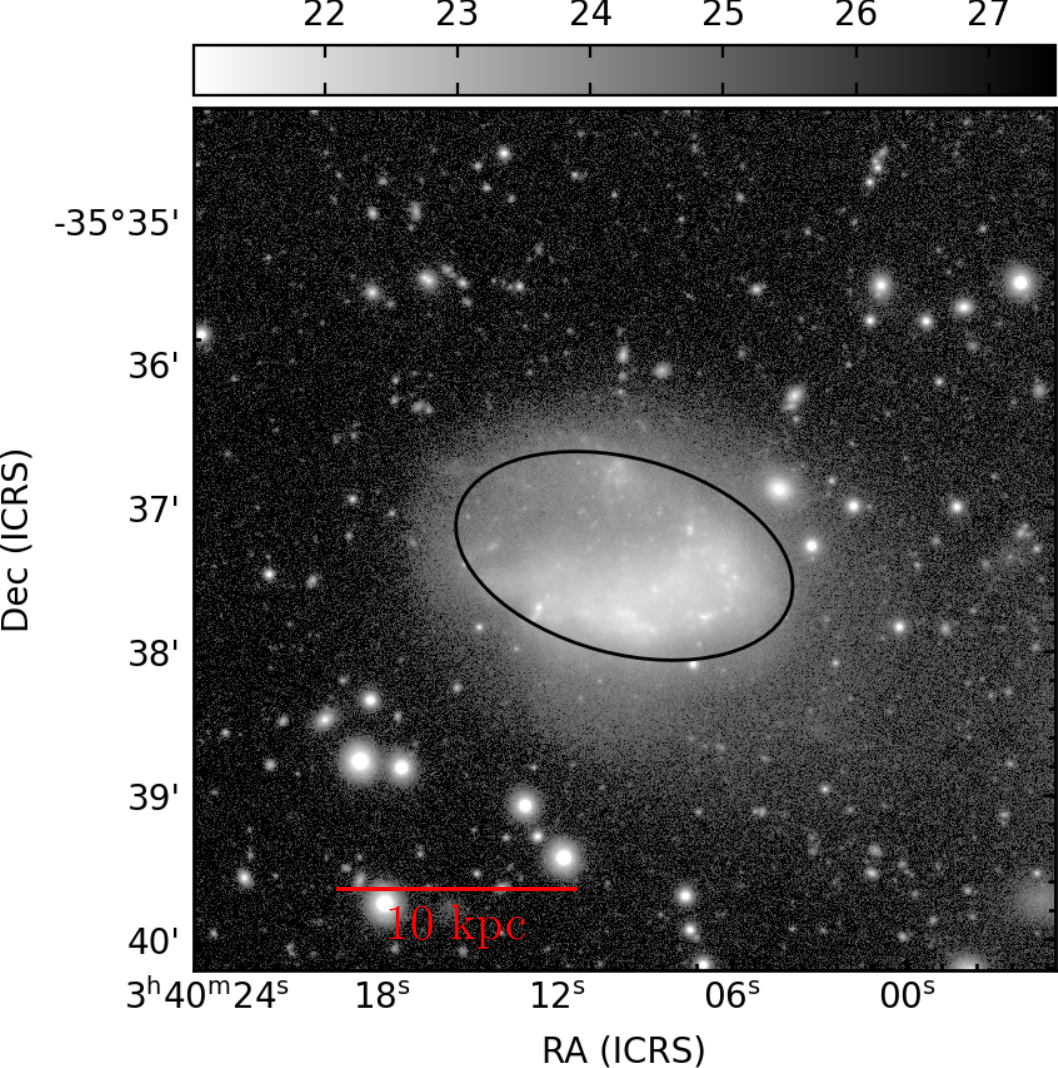}
  \caption{}
  \label{fig:r_band}
\end{subfigure}
\caption{
Images of NGC~1427A from \citet{Lee-Waddell2018} for illustrative purposes:~(a) VST image with overplotted in red contours of constant \Hi{} column density at levels $(0.5, 1, 2, 5, 10) \times 10^{20}$~amu~cm$^{-2}$;
(b) gas kinematics (recessional velocity in km/s)
(c) $r'$-band image (in mag/arcsec$^2$).
A length scale at the assumed cluster distance is indicated in red.}
\label{fig:NGC_OBSERVATIONS}
\end{figure*}
NGC 1427A is a gas-rich dwarf irregular galaxy in the Fornax Cluster.
Recessional velocity measurements indicate that NGC~1427A is moving with a line-of-sight velocity of around 2028 km/s \citep{Bureau1996, Schroder2001}.
Accordingly, if we assume NGC~1399 to be the cluster centre, NGC~1427A travels away from us at a projected speed of around 700~km/s w.r.t. the cluster centre. NGC 1427A has a projected distance of 137 kpc from NGC1399\footnote{The angular separation is 1373$\arcsec$ = 137~kpc, throughout the paper, we assume a fiducial distance of the cluster centre of 20~Mpc.}.

A false-colour HST image of the galaxy is shown in Figure~\ref{fig:NGC1427A}.
The inset of Figure \ref{fig:NGC1427A} zooms in on the Northern Clump, showing it to be composed of two loose stellar clusters with embedded H{\sc ii} regions. The Northern Clump also appears to be connected to the north-west rim of NGC~1427A's main body via a tenuous stream of stars. The directions of the \Hi{} tail and towards the Fornax Cluster centre are indicated with arrows. The dotted ellipse is the same as in Figures 1 and 2 of \citet{Lee-Waddell2018} and indicates the shape and direction of the faint outskirts of the galaxy. We interpret these low-surface-brightness stellar features, which point towards the north-east and south-west (effectively captured by the ellipses in Fig. \ref{fig:NGC1427A} and \ref{fig:r_band}), are quite distinct from the much brighter inner regions of the galaxy's main body, as tidal features. The Northern Clump, which we discuss in detail in section \ref{NC1427A}, may be associated with these tidal features.

\citet{Chaname2000} presents a lower limit for the dynamical mass of NGC 1427A using a rigid-body rotation model: $M_{\mathrm{dyn}}~>~(9\pm 3)\times 10^9$~\Msun{}. Following the calibrated empirical relation of \citet{Taylor2011}, the stellar mass estimated by \citet{Lee-Waddell2018} is $M_{\star} \approx 10^9$~\Msun{} while its total \Hi{} mass is determined to be $M_{\rm HI} = (2.1 \pm 0.2) \times 10^9$~\Msun{}.
The galaxy has a conspicuous \Hi{} tail, containing about $10$ per cent of all \Hi{} gas in NGC~1427A, pointing towards the south-east as well as stellar tidal extensions along a north-east to south-west axis. In other words:~the gaseous and stellar extensions are almost perpendicular to each other. NGC~1427A was not detected in CO emission, providing an upper limit on its molecular gas mass of the order of $M_{\rm H_{2}}~\sim 10^8$~\Msun{} \citep{Zabel2019}. These authors report the detection of a single 3~mm continuum source without an optical counterpart but its nature remains unclear.
A distinct young stellar clump is visible in the northern rim of the galaxy. On high-resolution images, this clump, the rim, as well as the galaxy's main stellar body are resolved into many individual OB associations and clusters, some with accompanying H$\alpha$-emission \citep{Hilker1997, Sivanandam2014}. Clearly, star formation proceeds in many dispersed small bursts.

Various hypotheses regarding the evolution of NGC 1427A, and especially its extraordinary configuration of tails, have been proposed. These include recent interactions with other nearby galaxies \citep{Cellone1997}, a merger with an object that now forms the northern stellar clump \citep{Lee-Waddell2018}, the tidal interaction with the cluster potential, and ram-pressure stripping of its gas by the cluster hot gas halo \citep{Chaname2000, Mora2015}.

To try and identify the environmental processes acting on NGC~1427A, we first selected those properties of this dwarf galaxy that were most likely to be indicative of those processes and not of internal effects. Those are:
\begin{enumerate}
    \item the direction towards the Fornax cluster centrum,
    \item the direction of the axis of the stellar tidal extension,
    \item the direction of the \Hi{} tail.
\end{enumerate}
These parameters are most tightly linked with NGC~1427A's orbit through the Fornax Cluster and are expected to be relatively insensitive to the accidents and vagaries of its evolution before its acquisition by the cluster.

Thus, our analysis differs from that of \citet{Lee-Waddell2018} where the ram-pressure hypothesis is discussed based on stellar colour and galaxy morphology information.
Their new data indeed rule out a ram pressure origin for the optical appearance of the galaxy, but they leave open the question whether ram pressure may have played a role in the formation of the \Hi{} tail.

\section{Simulations} \label{sec:simulations}
\begin{figure*}
\begin{subfigure}[b]{0.32\textwidth}
 \centering
 \includegraphics[width=\textwidth]{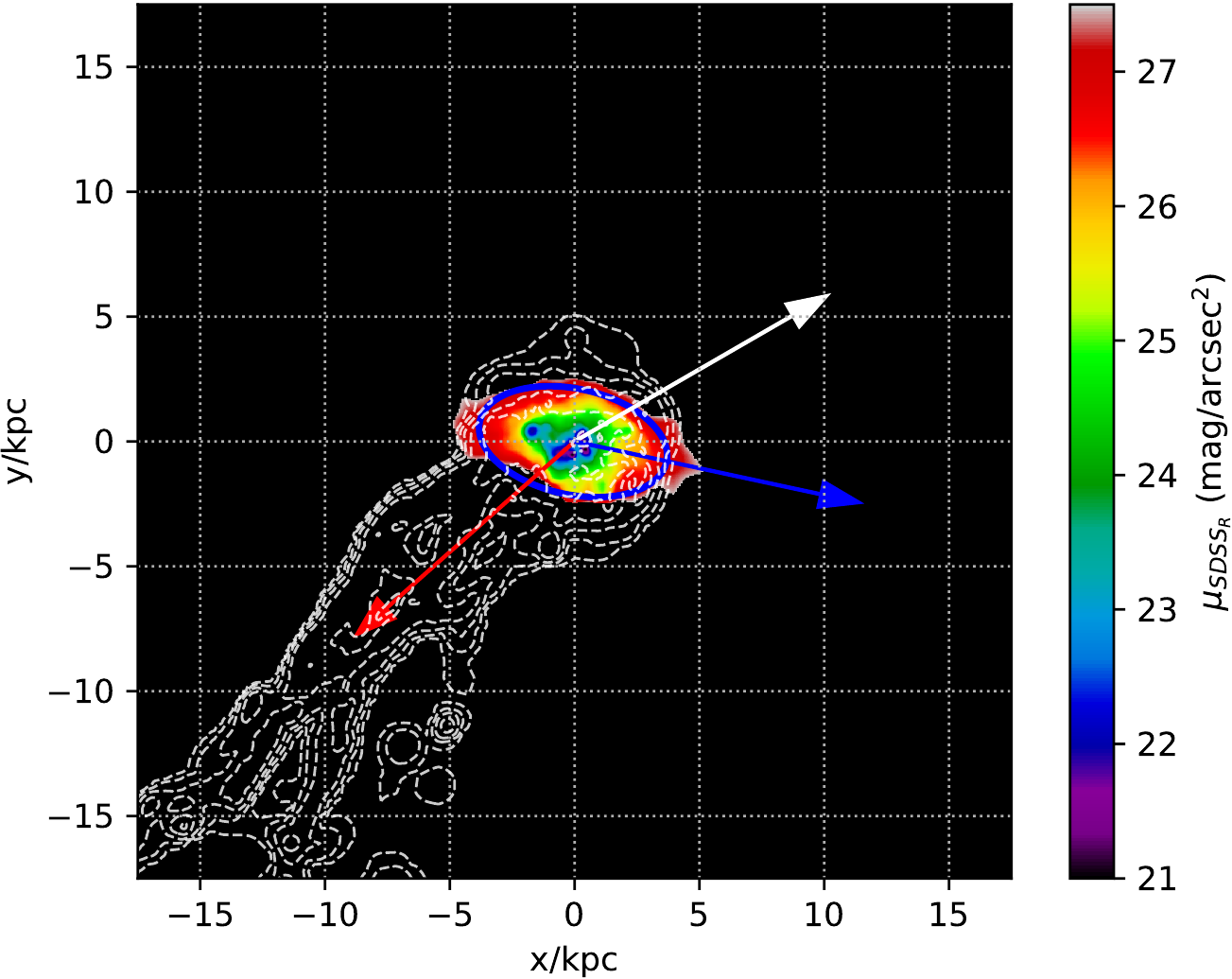}
 \caption{}
 \label{fig:sb_arrows}
\end{subfigure}
 \hfill
  \begin{subfigure}[b]{0.33\textwidth}
  \includegraphics[width=\textwidth]{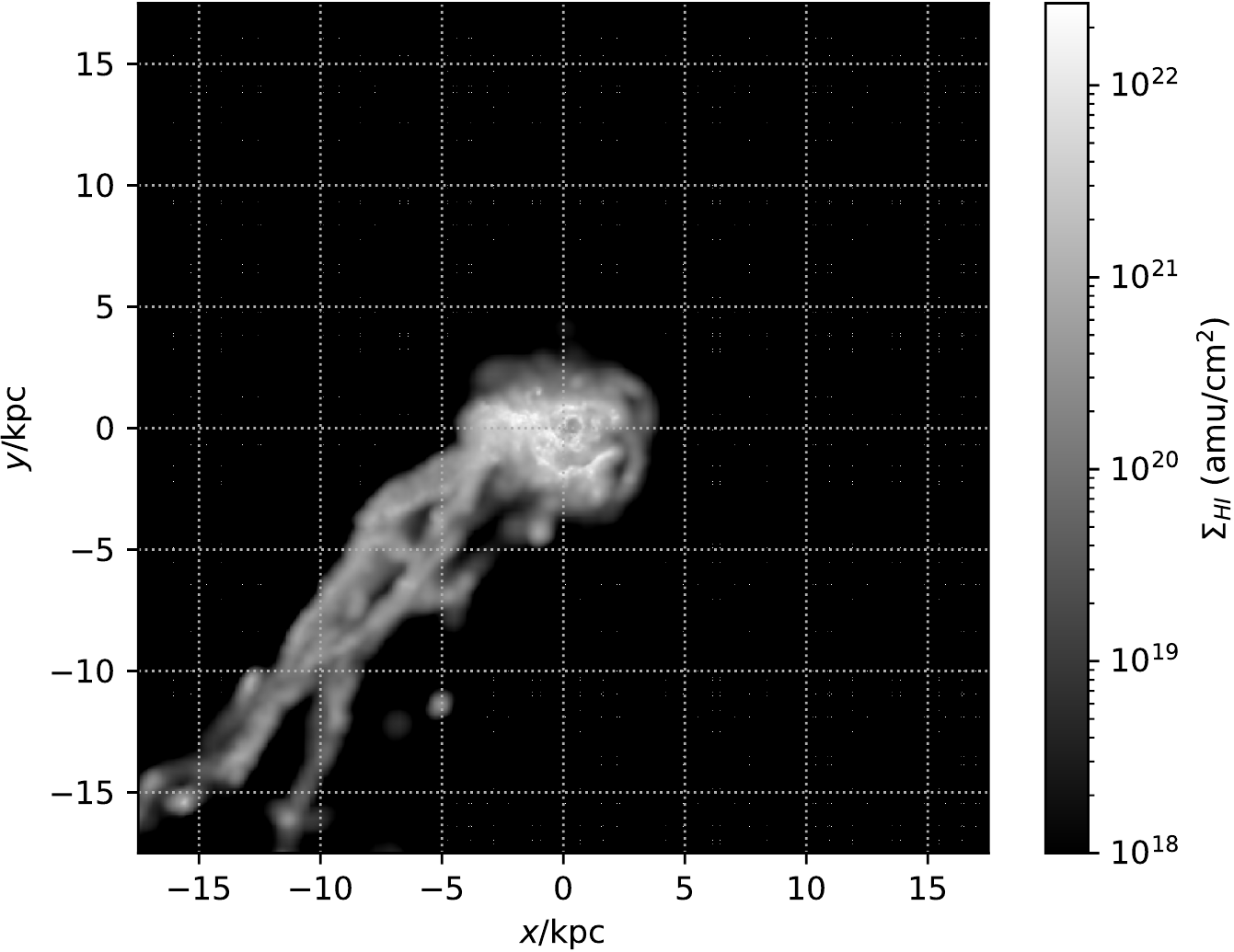}
  \caption{}
  \label{fig:sim_hi_density}
\end{subfigure}
\hfill
\begin{subfigure}[b]{0.33\textwidth}
  \centering
  \includegraphics[width=\textwidth]{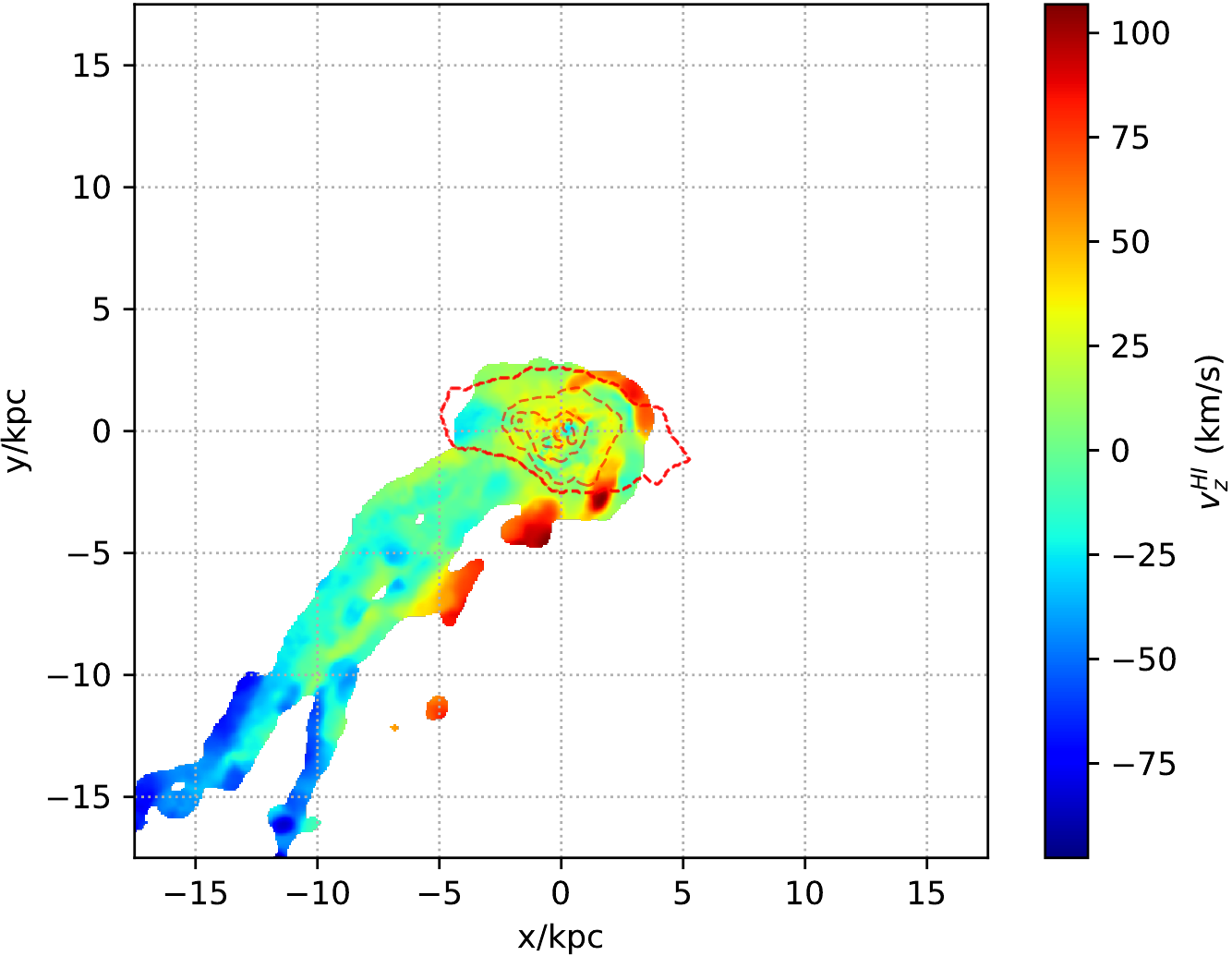}
  \caption{}
  \label{fig:sim_hi_kin}
\end{subfigure}
 \vfill
\caption{A simulation snapshot (ID 68, see Table \ref{tbl:sim}) on an orbit with 100 kpc pericenter distance, showing stellar tidal and gaseous stripped tails.
(a) $r'$-band surface brightness with \Hi{} column density contours as seen from a point of view for which the projected cluster-centric distance $r_p = 77$ kpc and the recessional velocity $-793$ km/s.
\Hi{} contours are drawn at column density levels [$10^{17}, 10^{18}, 10^{19}, 10^{20}, 10^{21}$] amu cm$^{-2}$. The white arrow points towards the cluster centre, whereas the blue and the red arrows are the computed orientation of the stellar tail and the gaseous tail respectively, see Section \ref{sec:morphological_quest}.
(b) The column density of the neutral hydrogen, highlighting clumpy blobs of stripped \Hi{}.
(c) SPH map of the \Hi{} velocity field where the \Hi{} tail in its rightmost part is receding more than the more detached part.
For reference, the [22, 24, 26, 28] mag/arcsec$^2$ isophotes are shown as dashed red contours.
We stress that our Fornax cluster dwarf galaxy simulations were not designed to reproduce all the details of this particular galaxy. Nonetheless, remarkable similarities can be found.}

\label{fig:selected_snapshot}
\end{figure*}

Starting from the MoRIA \citep[Models of Realistic dwarfs In Action, ][]{Verbeke2015, Verbeke2017}
suite of simulations, we performed a set of simulations to study the evolution of late-type dwarf galaxies in a Fornax-like cluster environment using the moving box technique \citep{Nichols2015}.
MoRIA dwarfs are cosmologically motivated models of galaxies ranging from $10^{6.5}$~\Msun{} to $10^9$~\Msun{} in stellar mass.
Dwarfs are evolved for around $8$~Gyr (up to $z=0.5$) using the merger-tree technique \citep{Cloet-Osselaer2014}.
With this method, the resulting halo is obtained through a sequence of proto-galaxies merging events which follow a cosmologically motivated merger tree.
All the simulations are performed using a $N$-body/SPH code which has been implemented on top of the publicly available code GADGET-2 \citep{Springel2005}.
Additional prescriptions of astrophysical processes have been added: star formation, radiative cooling, supernova feedback, UV background effects, and ionisation-aware internal energy and equation of state of the gas \citep{DeRijcke2013, Vandenbroucke2016a}.
The simulations can reproduce a whole range of observational properties of dwarfs in the field \citep{Schroyen2011, Verbeke2017}.

Using SPH it is computationally challenging to simulate an entire cluster of hot gas while at the same time having the resolution to properly treat the interactions at the interface between the interstellar and intra-cluster medium (or ICM) that cause ram pressure stripping.

We have opted to use the moving-box technique described by \citet{Nichols2015} and further developed by \citet{Hausammann2019}.
We enclose the MoRIA dwarf in a 60~kpc wide moving simulation box, as in a wind tunnel simulation.
Gas is injected from the open "front" side of the box, which always points in the dwarf's direction of motion. Its density and temperature vary with position, as discussed below in paragraph \ref{sec:ICM}. This mimics the hot wind of the cluster halo gas as it streams past the orbiting dwarf galaxy.
Also, additional fictitious forces on the particles are included to take into account the rotation and orbital motion of the moving box.
This allows us to simulate the combined effects of tidal forces and ram-pressure stripping \citep[as studied by][]{Mayer2006} which are acting simultaneously on the dwarf without the necessity of simulating a galaxy cluster worth of intra-cluster gas.
A slight improvement in our implementation of the moving-box method is the use of a critically damped oscillator for the \emph{ad hoc} acceleration to keep the galaxy close to the centre of the box.
A typical simulation snapshot, such as the one shown in Figure \ref{fig:selected_snapshot}, contains around 150k gas particles and 16k star particles, each with a mass of 4000~\Msun{}.

We stress that the simulations used in this paper are part of a larger suite and have not been set up to mimic NGC 1427A in any particular way. Therefore, not all details can be expected to exactly match with observations.
Nonetheless, these simulations provide valuable insights not only into the phenomena at play but, more importantly, they can be used to infer the most likely current orbital phase of the galaxy.

\subsection{Cluster model}

\subsubsection{Dark matter}
The simulations take into account both ram pressure stripping and the tidal interaction with the cluster.
The latter is simulated as a single spherically symmetric static NFW potential profile \citep{Navarro1996} with mass $M~=~10^{14}$~\Msun{} \citep{Drinkwater2001a}:
\begin{equation}
    \Phi(r) = \frac{G M}{r} \frac{\log(1+r/R_s)}{\log(1 + c) - \frac{c}{1+c}}
\end{equation}
with scale length $R_s = 120$ kpc and $c=8.15$ derived from scaling relations in e.g. \citet{Gentile2004, Wechsler2002}:
\begin{equation}
    c \simeq 20 \left(\frac{M}{10^{11} \text{\Msun{}} }\right)^{-0.13}, \qquad
    R_s = \frac 1 c \left(\frac{M}{\frac 4 3 \pi 200 \rho_c}\right)^\frac 1 3,
\end{equation}
with $\rho_c = 127.3$~\Msun{}~kpc$^{-3}$ the critical density of the universe at $z=0$ for a cosmology with Hubble constant $h=0.67$ and $\Omega_m = 0.31$ \citep{Planck2015}.
The cluster virial radius of the model is $R_{\mathrm{vir}} = c R_s = 978$~kpc.

\subsubsection{ICM} \label{sec:ICM}
\begin{figure}
\centering
\includegraphics[width=.9\columnwidth]{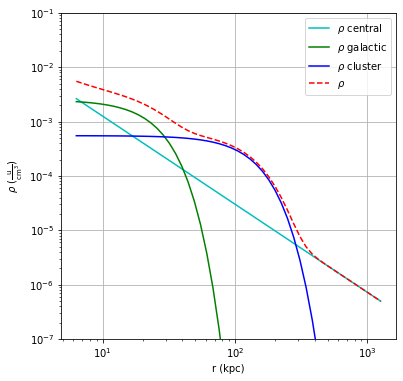}
\caption{Gas density profile $\rho(r)$ of the  Fornax Cluster model as a superposition of three beta models from measurements described in \citet{Paolillo2002}, from which we also adopt the nomenclature for the different beta models.
}
\label{fig:profiles}
\end{figure}

Following \citet{Paolillo2002}, we use the superposition of three spherically symmetric beta-models, $\rho(r) = \rho_0 (1 + (r/r_0)^2 )^{-3\beta/2}$, to construct the gas density profile in the Fornax cluster, as shown in Figure \ref{fig:profiles}.
They identify three contributions to the hot gas distributions: a central component (dominating for $r<5''$), coincident with the optical galaxy NGC1399; a less dense and more extended galactic component ($50''<r<400''$); and a cluster component ($r>400''$).
We assume the gas to be in hydrostatic equilibrium with temperature $T(r)$ computed as:
\begin{equation}
   T(r) = \frac {m_p}{k_B \rho(r)} \int_r^\infty \rho(r') \frac{GM(r')}{r'^2} \mathrm{d} r'
\end{equation}
where $M(r)$ is the mass of the gas, stars and dark matter beyond radius $r$, $m_p$ is the proton mass, $G$ and $k_B$ the gravitational and Boltzmann constants.

These two radial profiles allow us to inject particles with the proper density and temperature in the moving box, hence recreating the environmental condition of the cluster as the dwarf orbits through it.

\subsection{Simulation parameters}
We carried out a set of simulations starting from five MoRIA models of late-type galaxies taken at $z = 0.5$.
The overall goal in setting up the simulations is to study the evolution of late-type galaxies in a cluster environment.
The choice of the redshift of infall has been motivated by the fact that the number of red dwarfs in the Fornax cluster has increased significantly since $z = 0.5$ \citep{Stott2007, DeRijcke2010}.
That indicates that the conversion of late-type to early-type dwarfs (and hence the acquisition of late-type dwarfs by clusters) is probably a recent event.

We selected dwarfs models covering the stellar mass range of $10^{7.5}-10^{9}$ and we injected each of them on 5 different orbits with pericenter distances of $50, 100, 150, 200,$ and $300$~kpc and with a fixed apocenter of 800 kpc.
The starting point of the infall is always at a radial distance of 600~kpc.
We chose a lower radial distance of the starting position w.r.t. the virial radius given the very low cluster density in that region and the low orbital velocity of the dwarf near apocentre.
In this way, simulations could be concentrated on the infall and the pericenter passages of the simulated dwarfs.
We evolved the galaxies for $5.5$~Gyrs up to $z=0$.
The initial stellar masses are reported in Table \ref{tbl:sim}. All simulations presented in this paper, at time of injection have exponentially declining SB profiles with Sersic index around 1.0.
Every $10$~Myr a snapshot is saved, yielding around 560 snapshots for each simulation.
This high snapshot cadence has proven to be important for the following analysis (see Section \ref{sec:morphological_quest}).
Adhering to the simulation goal of following the evolution of a gas-rich late-type dwarf in a Fornax-like cluster, the initial snapshot of the most massive dwarf (ID 41) has been taken at $z=0.4$ because at $z=0.5$ it was still undergoing a major merger event.
This is equivalent to having this galaxy falling into the cluster more recently.
No sizeable effect has been noted in simulations results highlighting a different behaviour with respect to the other galaxies \citep[which underwent their last merger before infalling to the cluster at $z=0.5$, see e.g.][]{Cloet-Osselaer2014}.
This has had the only implication of a lower number of snapshots used in the technique explained in Section \ref{sec:morphological_quest}, but, as we shall see in the following, given that first pericenter passage turns out to be the most significant orbital phase, no notable bias is expected.

\begin{table}
	\centering
	\footnotesize
\begin{tabular}{cx{1.3cm}x{0.5cm}x{0.8cm}x{0.7cm}x{0.4cm}x{1.4cm}}
\toprule
Sim ID & $\log_{10}$(M$_\star$)\newline(M$_\odot$) & $R_e$ \newline (kpc) & $\sigma_\star$ \newline (km/s)\\
\midrule
  71 &  7.58 &  1.9 &  21.9 \\
  68 &  7.96 &  2.6 &  15.6 \\
  69 &  8.04 &  2.3 &  24.6 \\
  41 &  8.78 &  1.7 &  30.4 \\
\bottomrule
\end{tabular}
	\caption{Features at time of infall ($z=0.5$) of the selected MoRIA dwarf models used in this work}
\label{tbl:sim}
\end{table}

\section{Constraining the orbital phase of NGC~1427A} \label{sec:results}

\subsection{Quantitative morphological search}
\label{sec:morphological_quest}

\begin{figure}
\centering
\includegraphics[width=\columnwidth]{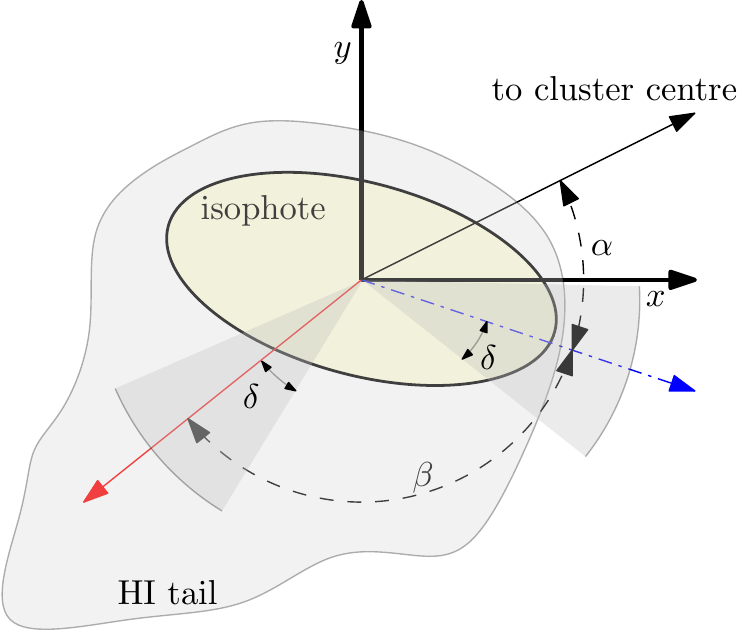}
\caption{Angles $\alpha, \beta$ definitions used to set up the  requirements for the morphological search. $\delta$ is the angle tolerance used for the selection. Blue and red arrows are defined as in Figure \ref{fig:sb_arrows}. Darker grey shaded regions are the allowed directions of the oriented snapshots tidal stellar tail and \Hi{} tail.}
\label{fig:angles_scheme}
\end{figure}

We carried out a systematic search among all the simulation snapshots, portraying different dwarfs at different times on different orbits.
We observed each snapshot from different points of view in order to select the snapshots and their orientation most resembling the observed galaxy using four measurable quantities:
\begin{enumerate}
    \item the projected cluster-centric distance, $r_p$;
    \item the line-of-sight velocity with respect to the cluster centre,~$v_p$;
    \item the position angle of the outer isophotes, quantified as the angle $\alpha$ between the projected cluster centre direction and the direction of the $26.5$~mag/arcsec$^2$ isophote in $r'$ band, see Figure~\ref{fig:angles_scheme};
    \item the orientation of the \Hi{} tail as measured with the angle $\beta$ between the isophote orientation and the direction of the highest variance of the image obtained by computing the second order moments of the \Hi{} SPH map \citep{Stobie1980}.
\end{enumerate}
We highlight that the search has been carried out using only morphological and orbital parameters, given our focus on the peculiar morphology of NGC~1427A.
Moreover, the criteria used to compare simulations to observations are the ones that best represent the effects of the cluster environment: the relative directions of the tidal pull and the ram pressure stripping. Indeed, the first two requirements put constraints on the orbital position and hence on the orbital phase of the simulated dwarf. The latter two require that the isophotal tail and the neutral gas tail are oriented as in NGC~1427A.
By selecting a faint isophote for criterion (iii) we are tracing the outer tidal extensions of the stellar body.
Inspection of the evolution of the direction of the principal axes of the stellar density distribution along a simulated dwarf galaxy's orbit shows that the galaxy outskirts rapidly forget their initial spatial orientation and become strongly aligned with the cluster centre around pericenter and apocenter. On the inward or outbound legs of a galaxy's orbit, this elongation occurs parallel with its orbital velocity. This evidence for a tight correlation between a dwarf galaxy's elongation and its orbital phase, led us to include criterion (iii).

From VST images of \citet[][]{Lee-Waddell2018} \citep[originally from the deep survey presented in][]{Iodice2016} we used the position angle of 15~deg of the fitted ellipse in its Figure 2.
From the \Hi{} maps of Figures 1 and 3 of the same paper (see Figures \ref{fig:hi_contours} and \ref{fig:hi_kin} above) we assumed an inclination of the \Hi{} tail of -135~deg (pointing south-east).
Given that the direction of the cluster centre is around 30~deg north-west, the resulting target angles are $\bar \alpha = 45$~deg, $\bar \beta = 120$ deg, as in Figure~\ref{fig:angles_scheme}.

\begin{figure}
\includegraphics[width=\columnwidth]{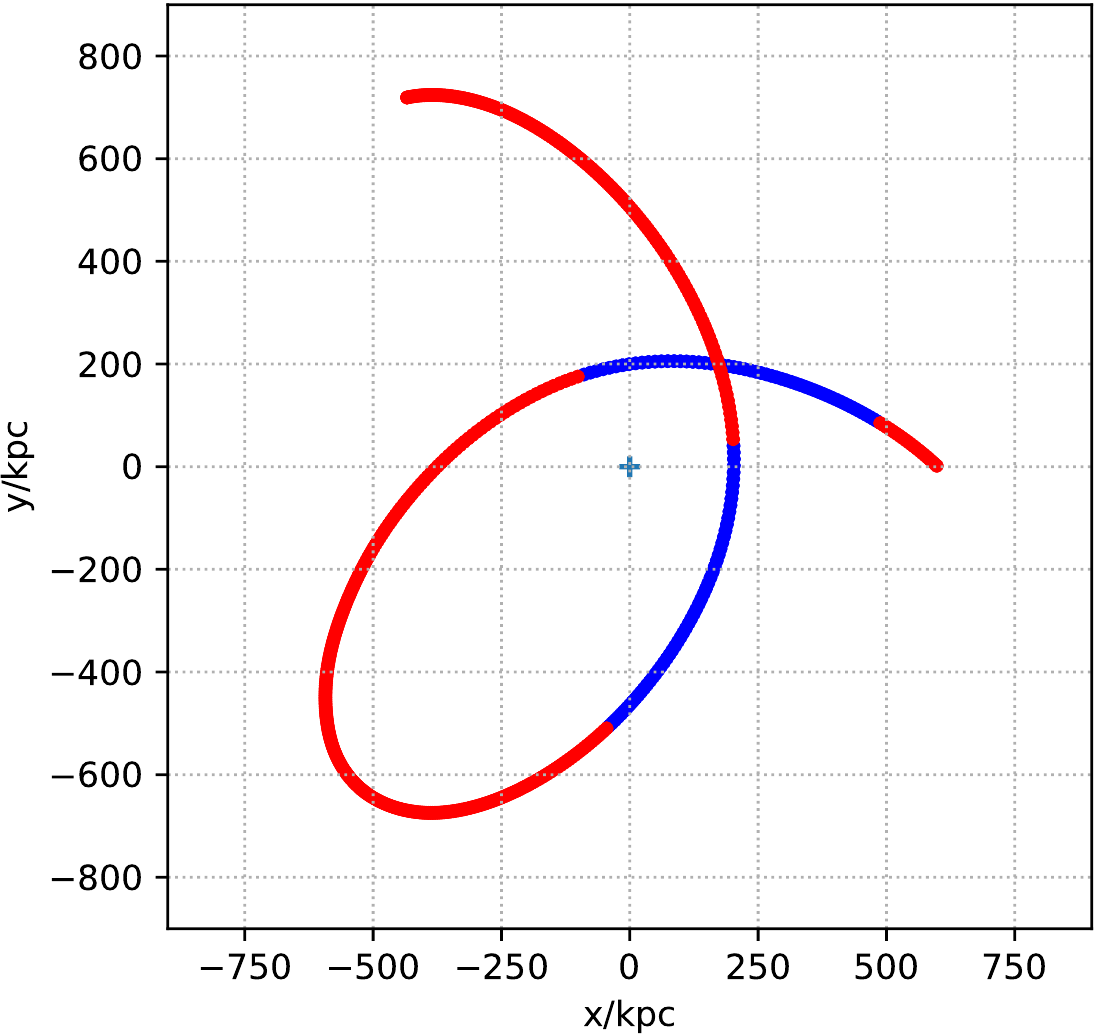}
\caption{A trajectory with pericenter of 200~kpc. In blue the orbital phases for which there are snapshots with projected distance and recessional velocity compatible with the ones of NGC~1427A (i.e. snapshots fulfilling requirements (i)-(ii)) for some points of view.
In this case $r_p = 137$~kpc, $v_p = -693$~km/s.
}
\label{fig:good_traj}
\end{figure}

For each snapshot, we started by finding the orbital phase which fulfils the first two requirements.
Each requirement is satisfied by the set of points of view constituting the generatrices of a cone centred on the galaxy position.
By intersecting two cones it is possible to find the points of view satisfying the requirements.
This is equivalent to solving a quadratic equation (see appendix \ref{sec:cone_intersection}) whose two solutions are two points of view satisfying the requirements (i) and (ii).
Each snapshot is then rotated as if it was observed from the peculiar point of view yielding the imposed clustercentric distance and the recessional velocity.
The more radial the orbit is, the higher the number of suitable snapshots which will be further selected using requirements (iii) and (iv), see Figure \ref{fig:good_traj}.

With the assumed cluster centre at a distance of 20~Mpc from us (see Section \ref{sec:intro}), we imposed as targets the two quantities $\bar r_p=137$~kpc and $\bar v_p=-693$~km/s with measures for NGC 1427A (the projected distance on the sky and recessional velocity of NGC~1427A w.r.t. NGC~1399).
In order to take into account uncertainties in the measured $\bar r_p$ and $\bar v_p$ and to capture the sensitivity of the procedure to the selected projected distance and recessional velocity, we repeated the same procedure for each simulated snapshot but allowing for other slightly offset targets $r_p$ an $v_p$.
Practically, we fixed an offset in both quantities $r_p = \bar r_p \pm \Delta r$, $v_p = \bar v_p \pm \Delta v$ with $\Delta r = 100$ kpc, $\Delta v = 60$ km/s.
Also we added four other targets: $(\bar r_p \pm \Delta r, \bar v_p)$ and $(\bar r_p, \bar v_p \pm \Delta v)$ with same ($\Delta r, \Delta v$) as before.
Including the exact target $\bar r_p, \bar v_p$, at the end, we had nine targets to check for each snapshot.
In total, we obtained a dataset of 424,656 oriented snapshots.

For each snapshot surviving the selection of requirements (i) and (ii) and oriented so that $r_p$ and $v_p$ are the ones imposed, we created the surface brightness map and \Hi{} map.
We first fitted an ellipse to the contour corresponding to the $26.5$~mag/arcsec$^2$ isophote in $r'$ band.
We computed the second order moments of the \Hi{} map to obtain the direction of the neutral hydrogen tail.
Since we are interested in snapshots with an elongated \Hi{} tail in the South-East direction, we further selected only oriented snapshots with galactic projected velocity on the plane of the sky having a positive projection on the cluster centre direction. This removes false positives with a \Hi{} tail inclined with the proper angle but extending towards the cluster centre (from the image moments only the direction is returned, not the sense of elongation of the tail).
At the end of this pre-selection, we ended up with 59,896 oriented snapshots.

We then used requirements (iii) and (iv) to further refine the search.
In the following section we shall determine the distribution of the snapshots with tails similar to the observed galaxy using angles $\alpha$ and $\beta$, described above, and a tolerance $\delta$:
\begin{equation*}
\bar \alpha - \delta < \alpha < \bar \alpha + \delta, \qquad
\bar \beta - \delta < \beta < \bar \beta + \delta
\end{equation*}
The selection tolerance $\delta$ is then our main knob to filter-out snapshots oriented as NGC~1427A with angles $\bar \alpha$ and $\bar \beta$. The dependence of the results (and the number of oriented snapshots surviving the selection criteria (i-iv)) on its choice is quite strong, so in all the following histograms we highlight the $\delta$ chosen.

\subsection{Distribution along the orbit of points of view satisfying the requirements}
\begin{figure}
\includegraphics[width=\columnwidth]{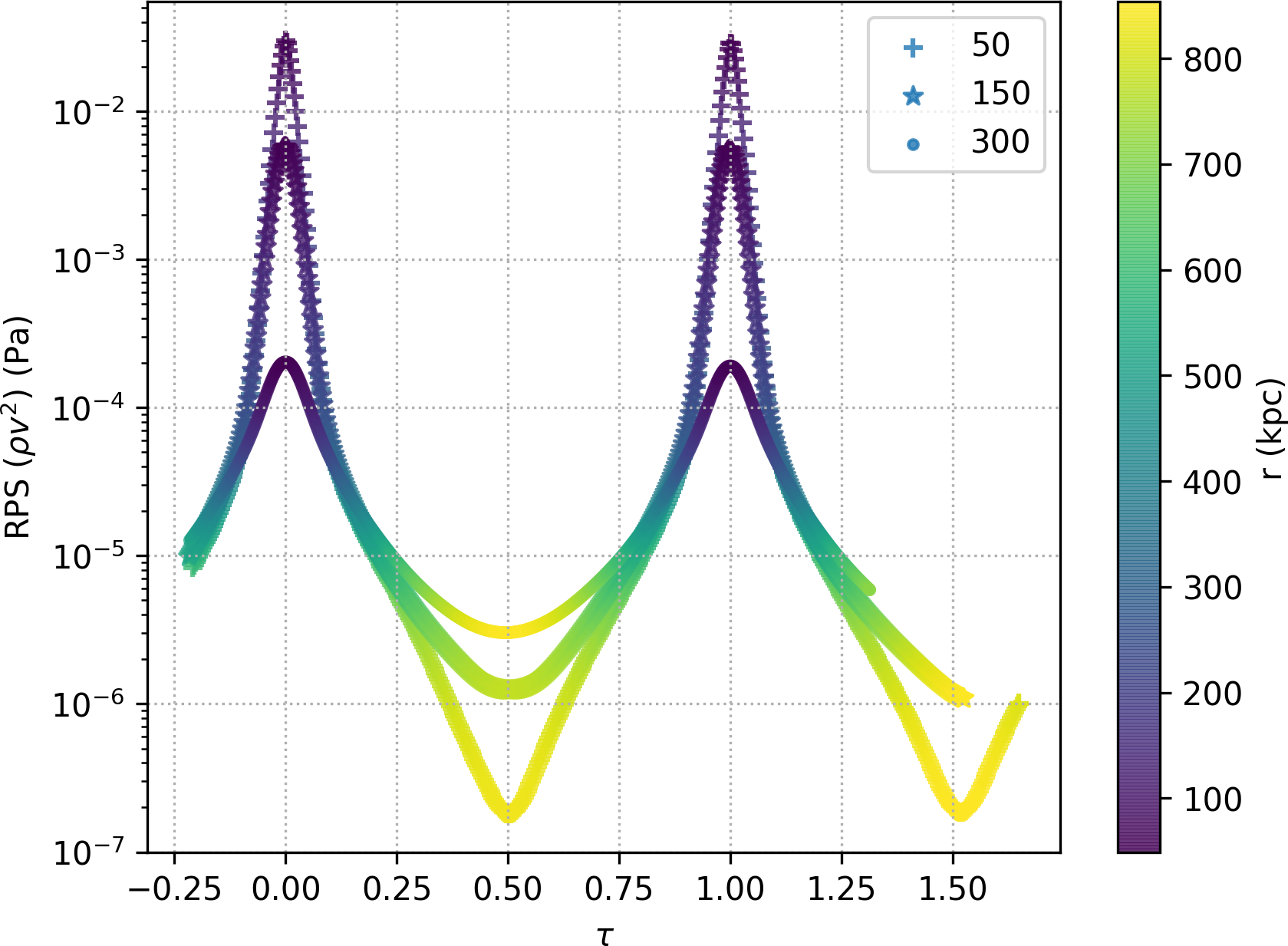}
\caption{The strength of the ram pressure ($\rho v^2$) as a function of normalised time $\tau = (t-t_p)/T_r$, where $t_p$ is the time of first pericenter passage and $T_r$ is the orbit's radial period, for the ID 69 simulations on multiple orbits (orbit pericenter in kpc is indicated in the legend). The colour scales traces orbital distance with respect to the cluster center.}
\label{fig:r_rps}
\end{figure}
We computed the distribution of selected oriented snapshots with respect to the time from the pericenter passage.
We noted that all the snapshots surviving the selection are found within 200 Myr of a pericenter passage.

\begin{figure*}
\includegraphics[width=0.8\textwidth]{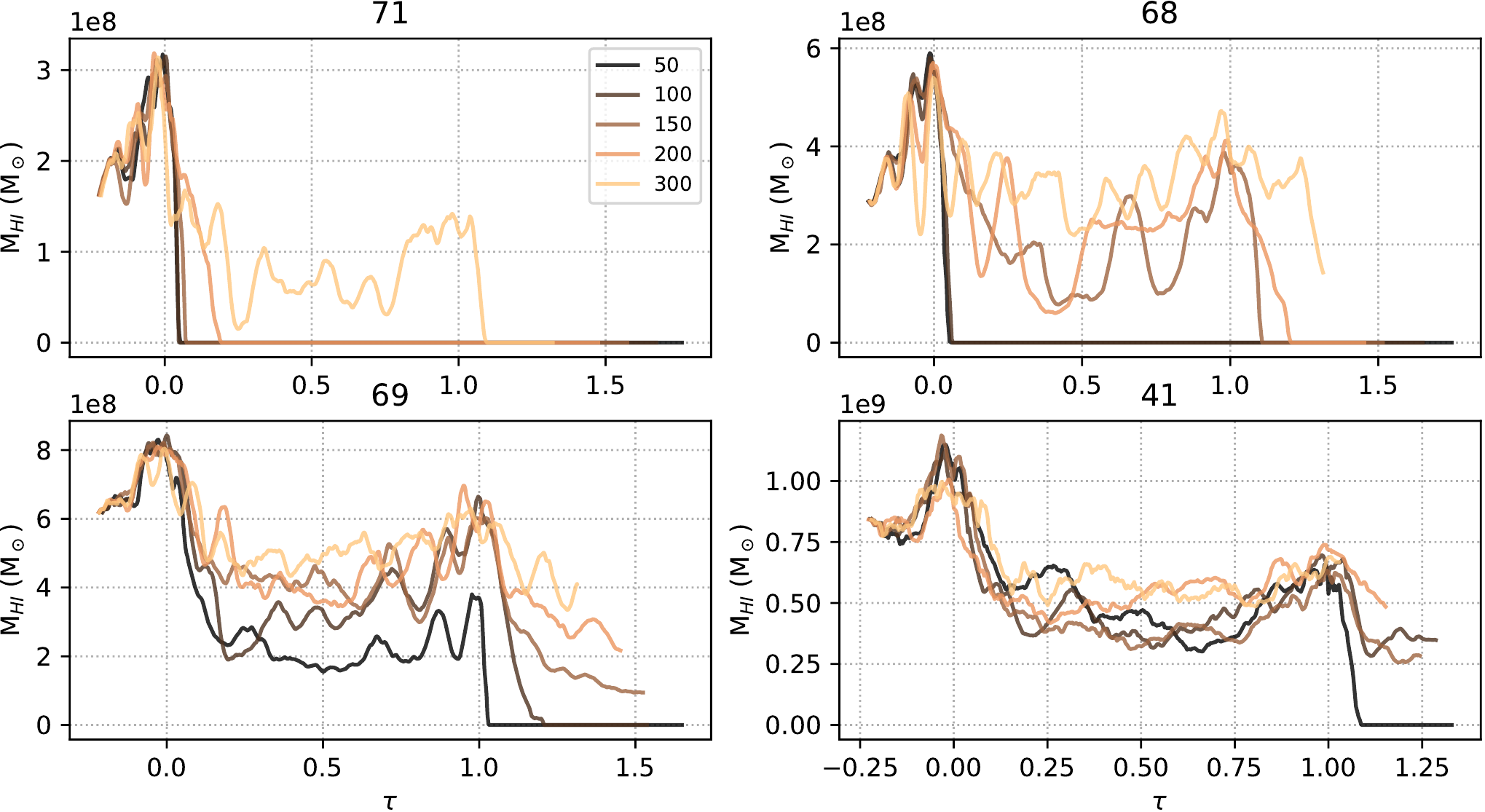}
\caption{Neutral gas mass as a function of normalised time $\tau$ as defined in Figure \ref{fig:r_rps}.
Each panel is labelled with the simulation ID (cf. Table~\ref{tbl:sim}) and contains information for the simulated dwarf launched on orbits with different pericenter distances (50, 100, 150, 200, 300 kpc identified by the colour legend in the top-left panel). Gas is compressed when the isolated galaxy enters the cluster and as a consequence it cools down, thus increasing the neutral hydrogen mass. Obviously, around pericenter, ram pressure stripping is effective at driving down the gas mass (as shown also in Figure~\ref{fig:r_rps}).
}
\label{fig:m_hi}
\end{figure*}

\begin{figure*}
\centering
\includegraphics[width=\textwidth]{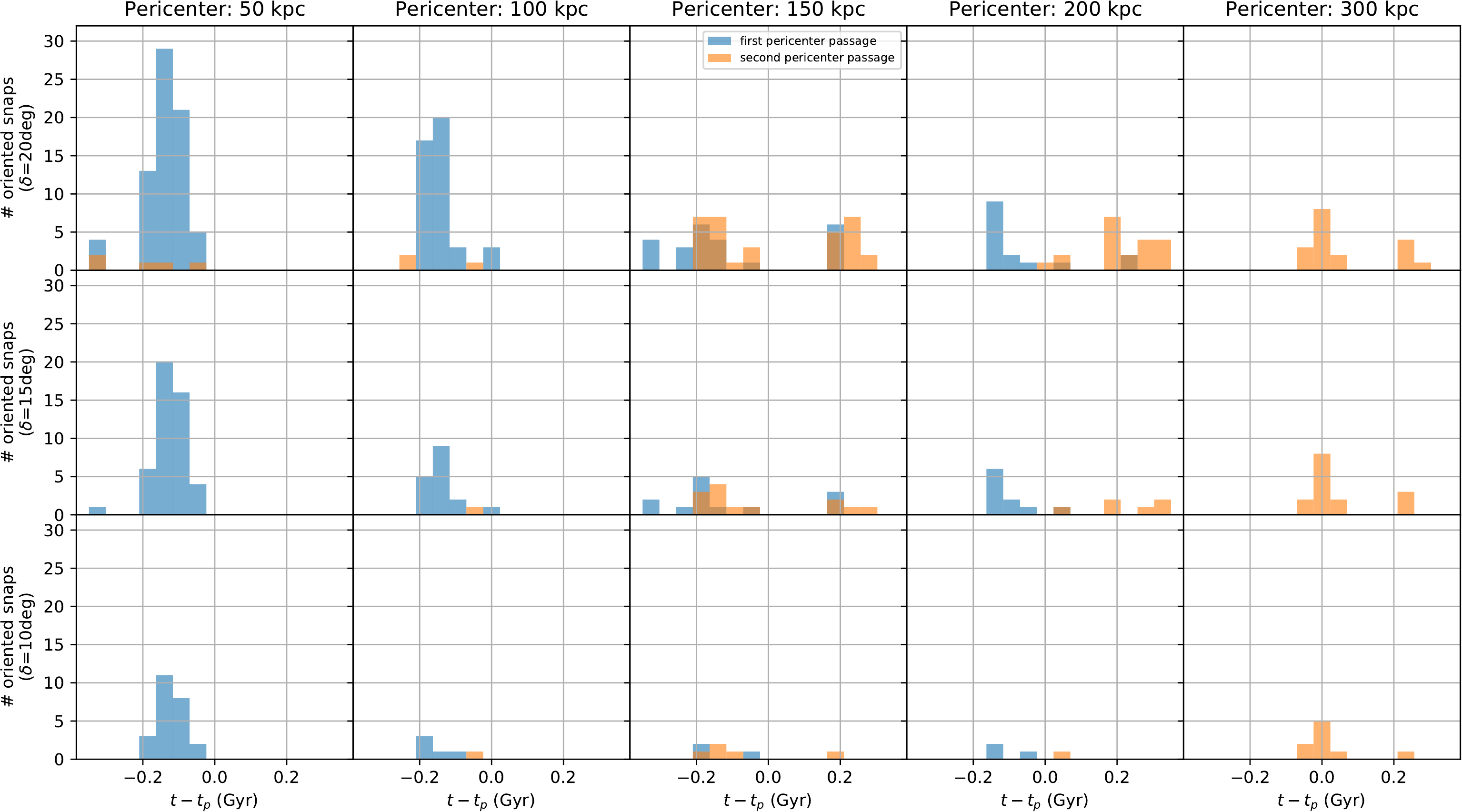}
\caption{Histograms of oriented snapshots fulfilling the requirements (i)-(iv), selected to be around first (blue) or second (orange) pericenter and grouped by orbital pericenter distances (50, 100, 150, 200, 300 kpc).
The isophote used to compute the stellar tail inclination is $26.5$~mag/arcsec$^2$ in $r'$ band.
The distribution is peaked at around $150$~Myr before pericenter passage, especially in more radial orbits. The result is robust enough to be visible on stricter angle tolerances $\delta = [20, 15, 10]$~deg.
}
\label{fig:histo_peri}
\end{figure*}

\begin{figure*}
\centering
\includegraphics[width=\textwidth]{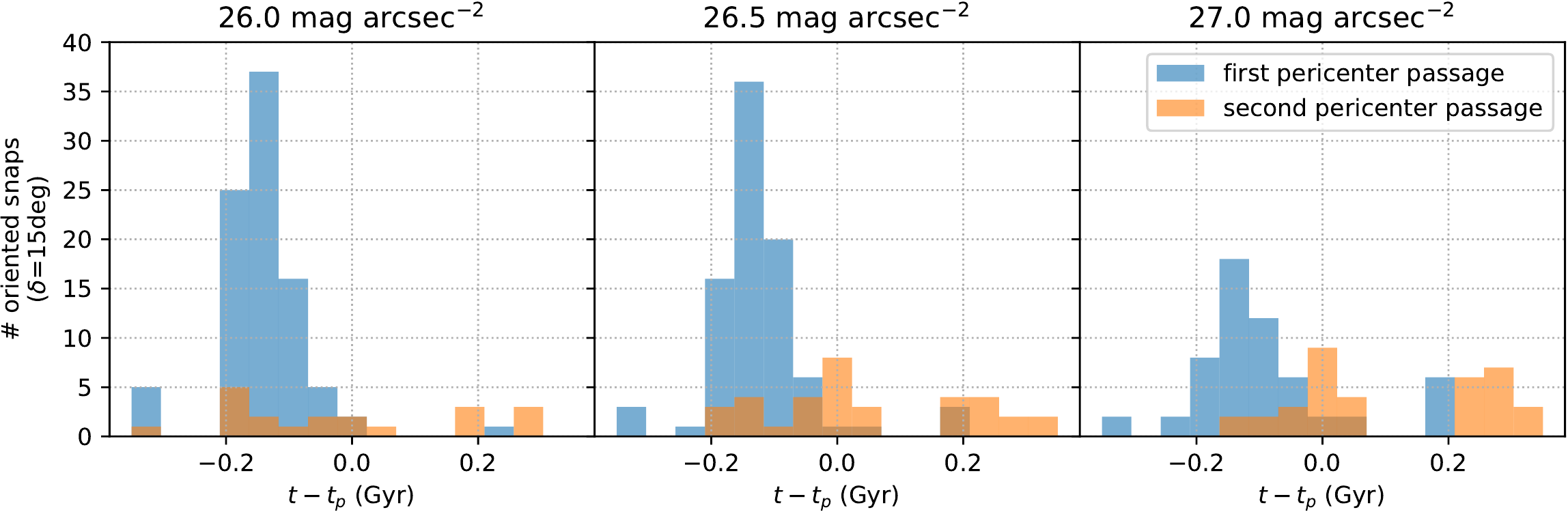}
\caption{Distribution in time relative to pericenter passage of oriented snapshots fulfilling the requirements (i)-(iv) with tolerance $\delta=15$ deg. Histograms of selected snapshots are coloured relative to their orbital phase: around first (blue) or second (orange) pericenter.
Each column corresponds to a different isophote used to compute the stellar tidal orientation.
Irrespective of the isophote, the distribution remains peaked  around $150$~Myr before pericenter passage.}
\label{fig:histo_noperi}
\end{figure*}

We ascertain the robustness of this result by varying the tolerance~$\delta$ of the comparison of the angles $\alpha, \beta$ with the observed ones.
Our orbital and morphological criteria are preferentially met by simulations with a stellar mass above $ \approx10^8 $ \Msun{}.
Less massive galaxies and galaxies on radial orbits are completely stripped from their HI gas, see Figure \ref{fig:m_hi}.
As shown in Figure~\ref{fig:histo_peri}, especially for radial orbits, no oriented snapshot is found on orbits of 50 and 100 kpc after first pericenter passage, regardless of the tolerance~$\delta$ considered.
For the most circular orbits, only snapshots undergoing second pericenter passage survive the selection. This is likely due to the first passage acting as 'preprocessing' and making the galaxy potential shallow with higher chances to create tidal tails.
In Figure~\ref{fig:histo_noperi} we performed the same analysis cumulatively counting all the oriented snapshots (without pericenter distance distinction) but using different isophotes.
The diagrams, even if noisier when using fainter isophotes, convey the same message: there is an abundance of correctly oriented snapshots (with tails as NGC~1427A) around first pericenter passage.
Indeed, tidal tails as those shown in Figure~\ref{fig:panel}, are present in low surface brightness regions of the dwarf, even if they are more difficult to measure.
The counterpart in NGC~1427A would be the faint stellar South-West elongation visible especially in $r'$-band, see Figure~\ref{fig:r_band}.

\subsection{3D position of the galaxy in the Fornax context}

Based on the above described model, we can produce a quantitative estimation of the galaxy projected radial distance. This measure is actually a testable prediction with distance observations.
Unfortunately, current uncertainties on the distance measurements do not allow to unequivocally assess the position of the galaxy to be in front or behind the cluster centre \citep{Georgiev2006}.

\begin{figure}
\centering
\includegraphics[width=\columnwidth]{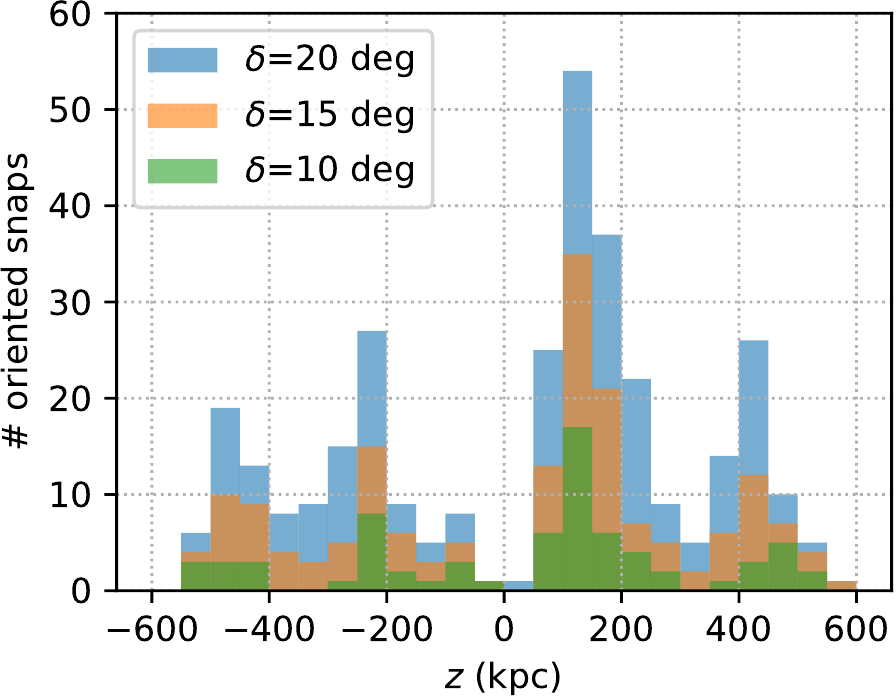}
\caption{Distribution of selected oriented snapshots with multiple tolerance $\delta$ along the projected line of sight. The zero point is 20~Mpc, assumed distance of NGC~1399. Positive values mean the galaxy being \emph{in front of} the cluster centre.
}
\label{fig:distance_prediction}
\end{figure}

Using our models we can see which is the most likely radial distance relative to the cluster centre of a galaxy with morphological features like NGC~1427A.
As shown in Figure \ref{fig:distance_prediction}, the preferred line of sight distance is around $200$~kpc in front of the cluster centre.

\begin{figure}
\centering
\includegraphics[width=\columnwidth]{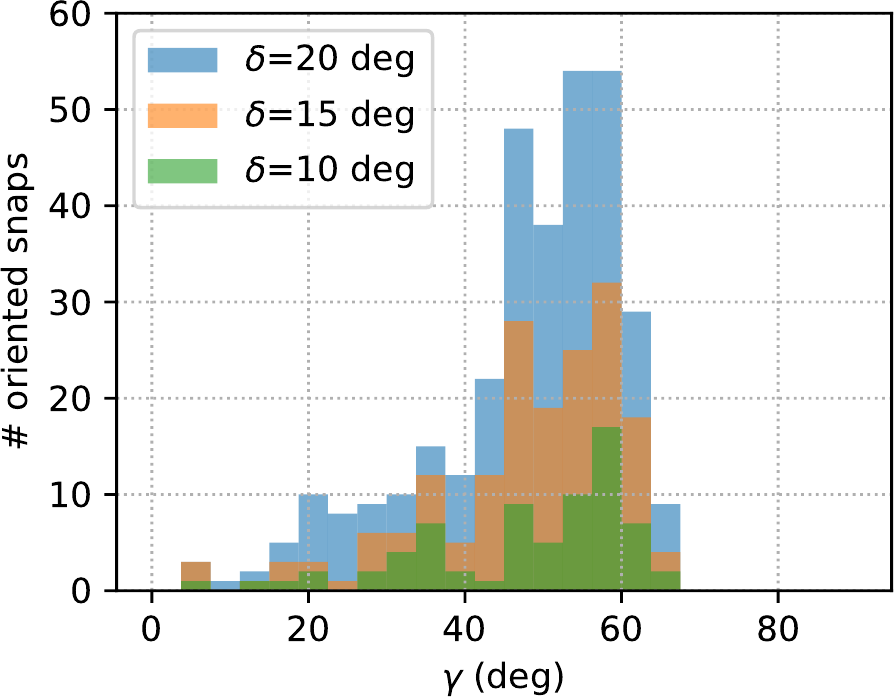}
\caption{Distribution of the flight angle $\gamma$ of selected oriented snapshots with multiple tolerances $\delta$. $\gamma$ is defined as the angle between the galactic velocity vector in the the cluster reference frame and the line of sight direction.
}
\label{fig:velocity_prediction}
\end{figure}

\begin{figure}
\centering
\includegraphics[width=0.6\columnwidth]{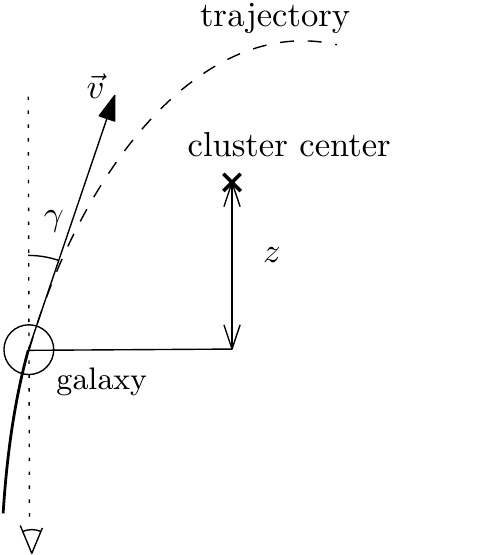}
\caption{Definitions of angle of flight $\gamma$ and radial distance $z$. $\vec v$ is the orbital velocity of the galaxy.
}
\label{fig:gamma}
\end{figure}

It is also possible to compute the flight angle $\gamma$ of the galaxy with respect to the line of sight direction (see Figure \ref{fig:gamma}).
As shown in Figure \ref{fig:velocity_prediction}, from the above models, the most frequent is $\gamma\approx 50$~deg.
Given that the stripped gaseous tail approximately follows the opposite direction of flight, measuring the flight angle in simulations can be useful to assess the real length of the gaseous tail in observations.
Our result would indicate the real tail to be roughly a factor 1.3 longer than the projected one.


\section{Discussion} \label{sec:discussion}

\subsection{The asymmetric stellar tidal tails}

\subsubsection{Including asymmetry as a constraint}
As a way of quantifying the asymmetric stellar tides, we computed the non-parametric measure Asymmetry \citep[as defined by][]{Lotz2004} of our oriented snapshots. For each oriented snapshot we fed its surface brightness map to \verb|statmorph| \citep{Rodriguez-Gomez2019} isolating the region of the map within $27$~mag/arcsec$^2$.
As a reference, \citet{Su2021} find an Asymmetry of $0.23$ for NGC~1427A.
We tried to add the Asymmetry to the constraints described in Section \ref{sec:morphological_quest}. Given that almost all the oriented snapshots have an Asymmetry higher than $0.2$ \citep[in line with the average of $0.53\pm0.22$ for galaxies with intense star formation as reported by][]{Conselice2003}, we find that isolating snapshots at least as asymmetric as NGC~1427A do not affect the results.

\begin{figure*}
\centering
\includegraphics[width=\textwidth]{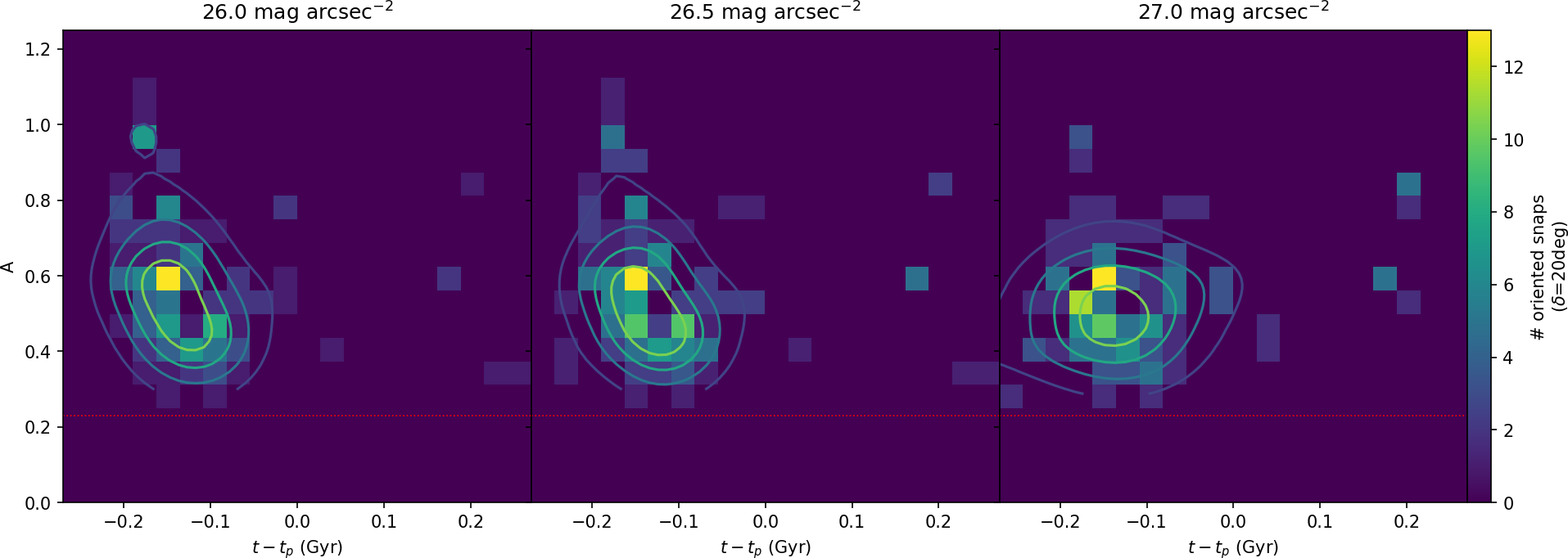}
\caption{2D histogram of selected oriented snapshots with tolerance $\delta=20$~deg.
The distribution is plotted with time from first pericenter passage on the $x$ axis, whereas on the $y$ axis the \emph{Asymmetry} non-parametric measure as defined by \citet{Lotz2004}.
Each column corresponds to a different isophote used to compute the stellar tidal orientation.
We overplot a kernel density estimation of the distribution for the snapshots approaching pericenter. The dotted red line corresponds to the measured \emph{Asymmetry} for NGC~1427A.}
\label{fig:histo2d_asym}
\end{figure*}

By plotting Asymmetry on selected snapshots as a function of time from pericenter, as shown in Figure \ref{fig:histo2d_asym}, it can be seen that selected snapshots closer to the pericenter become more symmetric.
A possible reason of this can be hypothesised in the tidal pull close to the pericenter which squeezes the galaxy elongating it, hence removing asymmetric regions of the galaxies.
Indeed all simulations show an Asymmetry greater than the one of NGC~1427A. We note that this is in line with \citet{Rodriguez-Gomez2019} who find a systematically higher asymmetry for simulated galaxy of Illustris and Illustris TNG \citep{Vogelsberger2014, Pillepich2018}, especially in the low mass range.
In fact, in a numerical simulation each star particle created represents stars of $\approx 1000$ \Msun{}: a simulated galaxy is always more "granular" than an observed real galaxy due to the limited resolution.

\subsubsection{The origin of the asymmetric stellar tidal tails}
\begin{figure*}
\includegraphics[width=\textwidth]{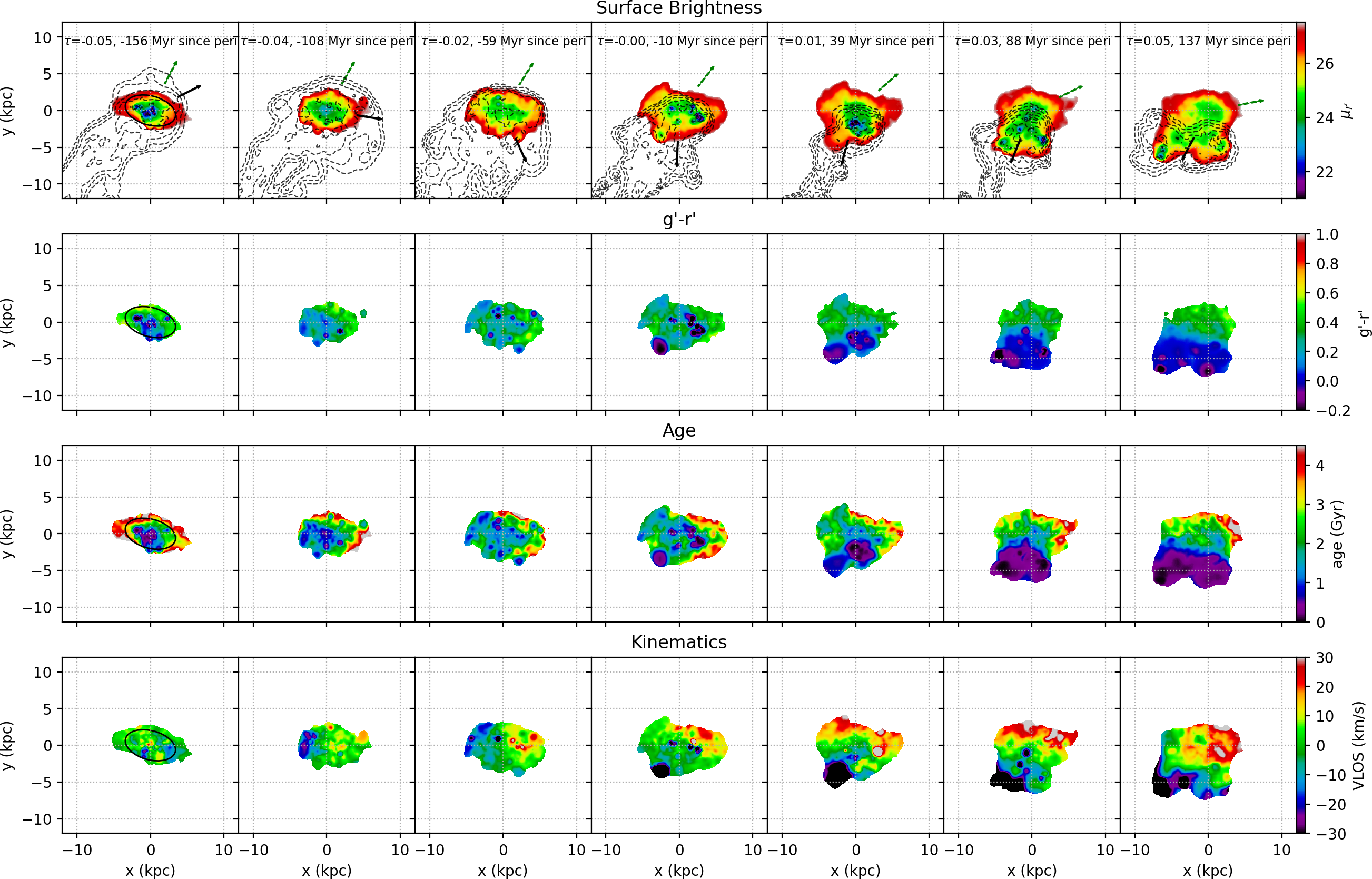}
\caption{Evolution in time of the simulation with ID 68 around first pericenter passage, projected in a way that the first column has $r_p = 137$~kpc, $v_p = 693$~km/s, the target values for NGC~1427A.
All the other snapshots are seen from the same point of view as the first one.
Black arrow indicates the direction to the cluster centre whereas the green indicates the instantaneous velocity direction of the galaxy projected on the plane of the sky. In the first column, an ellipse fitted on the $26.5$ mag/arcsec$^2$ is shown to highlight the direction of the tidal elongation.
The first row represents the surface brightness of snapshots around the first pericenter. \Hi{} Contours are [$10^{17}, 10^{18}, 10^{19}, 10^{20}, 10^{21}$]~amu~cm$^{-2}$.
Second row the $g'$-$r'$ colour.
Third and fourth row the v-band SPH-average age and velocity of star particles along the line of sight.}
\label{fig:panel}
\end{figure*}

An interesting effect of the pericenter passage is the formation of an asymmetric tidal tail, a stellar elongation more pronounced in just one direction, as that shown in the fourth column of Figure~\ref{fig:panel}.
This effect can be investigated by looking at different simulation snapshots evolving with time around a pericenter passage.
We focus our discussion to the case of the galaxy with ID 68 on an orbit with pericenter $100$~kpc.
In Figure~\ref{fig:panel}, the snapshot in the first column is the one who has passed the filters described in Section~\ref{sec:morphological_quest} and represents a snapshot which is in good agreement to the observation, given its stellar and gaseous tail directions and the projected clustercentric distance and recessional velocity.

We can then reconstruct a series of events leading to the generation of the leading edge stellar tail.
The tidal forces exerted by the Fornax Cluster become markedly asymmetric as a galaxy approaches pericentre on a radial orbit. The steeply deepening gravitational potential well can raise a stronger leading stellar tidal tail while producing a weaker trailing tail during a fast swing-by of a galaxy close to the cluster centre. After pericentre, the leading tail twists due to the curvature of the galaxy's orbit, the rapidly changing direction of the gravitational force, and the internal rotation of the galaxy.
Effectively the pericentre passage injects energy into the galaxy resulting in a temporary increase of its angular momentum.

At the same time, the gaseous tail is always directed opposite to the instantaneous velocity.
The result is that the stellar and gaseous tails are very misaligned, almost orthogonal to each other.

\subsection{Possible origin of the Northern Clump in NGC 1427A} \label{NC1427A}
NGC~1427A contains a so called Northern Clump (NC), which has been investigated in detail by some authors \citep{Cellone1997, Hilker1997}.
The NC is a clump of blue and very young stars with associated H$\alpha$ emission \citep{Sivanandam2014}. Even on ground-based images, the NC appears to be composed of two sub-clusters, one to the north of the other. This impression is strengthened by the HST image presented in Figure \ref{fig:NGC1427A}. The centres of the two sub-clusters are separated by 7\arcsec, corresponding to just under 700~parsec. The NC lies on NGC~1427A's projected rotation axis and hence its line-of-sight velocity agrees very well with NGC~1427A's mean recession velocity \citep{Bureau1996,Chaname2000}. This does not hint at an external origin for this object. The NC appears to be connected to the north-west rim of the galaxy's main body by a tenuous stream of stars, suggesting it to be displaced along this stream away from the galaxy's main body in a direction that is almost parallel to the major axis of the faint outer isophotes. If, as in our interpretation of the data, these outer isophotes trace the two diametrically opposite stellar tides raised by the Fornax Cluster forces then this would argue for a purely internal origin of the NC. It could be a star-formation region (of which there are many inside NGC~1427A) that is being pulled out by the Fornax Cluster tidal forces, leaving behind a stream of stars.

As shown in Figure \ref{fig:panel}, star formation flares up in our simulated dwarf galaxies around pericentre passage and leads to the appearance of scattered bright, blue clumps of active star formation. These clumps orbit along with the general rotation of their host galaxy. The spatial location and the time of appearance of these clumps are erratic and differ between galaxies and between orbits.
For instance, a star-forming clump is first visible to the top-right of the galaxy in the snapshots $108$~Myr before pericentre passage (second column in Figure \ref{fig:panel}; it rotates clockwise, and disappears again after the $137$~Myr past pericentre passage snapshot. Likewise, other clumps with similar lifespans pop in and out of existence around pericentre passage.

Based on these simulations and the available observational data, we suggest that the NC is precisely such a star-forming clump. This interpretation is consistent with its very young age and blue colour, its presence around the time that NGC~1427A is expected to be near pericentre passage (this is required to explain all other characteristics of NGC~1427A), and its kinematics being in line with the galaxy's global velocity field.

\section{Conclusions}
We carried out a set of simulations of gas-rich late-type dwarf galaxies in a Fornax-like cluster environment.
We isolated snapshots with morphological properties similar to the peculiar galaxy NGC~1427A. The properties have been chosen to be representing the impact of the environment on the dwarf.
We found that the main effects generating peculiar morphology are indeed the combination of ram pressure and tidal interaction close to the cluster centre and galaxy rotation.
We saw that the most likely scenario which recreates NGC~1427A tails morphology is assuming the galaxy to be on a very radial orbit with its tail almost aligned with the line of sight, pointing towards the observer.
This naturally leads to a gas kinematic configuration consistent with \Hi{} observations: in the westward part, gas attached to the stellar body of the galaxy is receding whereas the eastward part is stripped and dragged towards the observed by the intra-cluster medium (or ICM), therefore having a smaller recessional velocity, see Figures \ref{fig:sim_hi_kin} and \ref{fig:hi_kin}.

From the analysis of the morphology of the simulated snapshots motivated by environmental effects, we found an excess of snapshots revealing similar NGC~1427A structure around $150$~Myr before pericenter passage.
It should be highlighted that this result comes from a suite of simulation which has not been tailored from the beginning to reproduce NGC~1427A.
Nonetheless, interestingly, falsifiable predictions on the location and orbital phase of the galaxy can be made.

We can sum up the main results in the following points:
\begin{itemize}
    \item Perpendicular gaseous and stellar tails are explainable given that they are subject to different environmental effects;
    \item Tails geometry is crucial to unravelling the direction of motion of the galaxy and its orbit.
    \item From our suite of simulation it is evident how around \textasciitilde $150$~Myr before first pericenter passage, a morphological tail structure like the one of NGC~1427A emerges in galaxy falling into a Fornax-like cluster.
    \item In simulations, around pericenter, clumps of newly formed stars can form. This is coherent with a formation scenario of NGC~1427A's Northern Clump as a star formation region pulled out by tidal forces.
    \item Following our modelling it is possible to estimate the most likely position of a NGC~1427A-like galaxy to be around $200$~kpc in front of the cluster centre. Also, the most likely flight direction (represented by the angle $\gamma$ in the paper) is around 50 deg.
\end{itemize}

\section*{Acknowledgements}
We thank the anonymous referee for constructive suggestions that helped improve the content and structure of this paper.
We acknowledge financial support from the European Union's Horizon 2020 research and innovation programme under the Marie Sk\l odowska-Curie
grant agreement N.~721463 to the SUNDIAL ITN network.
M.M. thanks F. D'Eugenio and A. Nersesian for insightful discussions.
Analysis of data and plots have been made possible thanks to open-source software:
\verb|pynbody| \citep{Pontzen2013},
\verb|numpy| \citep{numpy},
\verb|scipy| \citep{scipy},
\verb|pandas| \citep{pandas},
\verb|matplotlib| \citep{Hunter2004},
\verb|astropy| \citep{TheAstropyCollaboration2018},
\verb|statmorph| \citep{Rodriguez-Gomez2019}.

\section*{Data availability}
The data underlying this article will be shared on reasonable request to the corresponding author.

\bibliographystyle{mnras}
\bibliography{biblio_short_journal}

\appendix
\section{Finding points of view - cones intersection}
\label{sec:cone_intersection}
Finding the point of views from which the galaxy appears as having the target projected clustercentric distance ($r_p$) and the proper line-of-sight velocity ($v_p$) is equivalent to solving the problem of intersecting two cones.

Given $\vec x$ the unit vector representing the direction of the point of view, and $\vec r$ and $\vec v$ the clustercentric position and velocity of the galaxy respectively, we can write the following conditions:

\begin{equation}
\label{eq:system}
\begin{cases}
\vec x \cdot \vec v = v_p\\
\vec x \cdot \vec r = \pm R\\
|\vec x|^2 = 1
\end{cases}
\end{equation}
where $R = \sqrt{r^2 -r_p^2}$. By using [$\vec r, \vec v, \vec r \times \vec v$] as basis (right-handed but not orthogonal), it is possible to express $\vec x$ as:
\begin{equation}
    \vec x = a \vec v + b \vec r + c ( \vec r \times \vec v).
\end{equation}
Substituting into \eqref{eq:system}:
\begin{equation}
\begin{cases}
a v^2 + b (\vec r \cdot \vec v) = v_p\\
a(\vec r \cdot \vec v) + b r^2 = \pm R\\
|\vec x|^2 = 1 = a v^2 + b r^2 + c^2 | \vec r \times \vec v|^2 + 2 ab(\vec r \cdot \vec v)
\end{cases}
\end{equation}
The last quadratic equation yields immediately two values of $c$ ($c_1$, $c_2$).

For each chosen sign of $R$, the system yields two solutions: ($\vec x_1$, $\vec x_2$) which can be used to rotate the galaxy snapshot as if it was seen from the directions $\vec x_1$ and  $\vec x_2$.

Each direction can be defined using two angles ($\varphi$ and $\theta$) representing the spherical coordinates of the unit vectors $\vec x_1$ and $\vec x_2$.

\begin{figure}
  \centering
  \includegraphics[height=6cm]{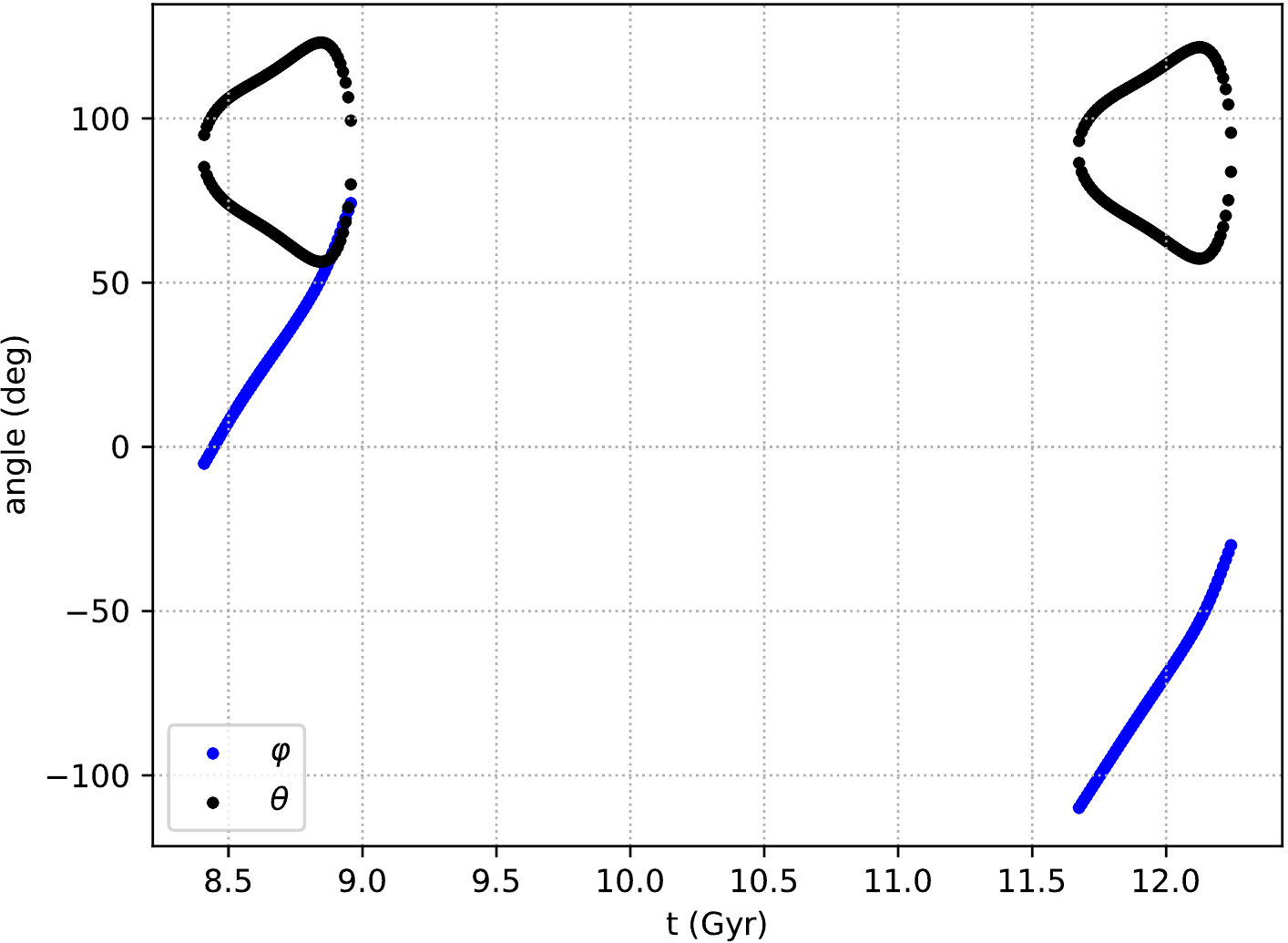}
  \caption{Angles $\varphi, \theta$ defining the points of view of each snapshot of the trajectory of Figure \ref{fig:good_traj}. The angles are used to rotate the selected snapshots in order to obtain the target $r_p, v_p$. Only a subset of snapshots can be oriented to fulfil the requirements.}
  \label{fig:phitheta}
\end{figure}
In Figure \ref{fig:phitheta} an example of rotation angles for a particular simulated orbit is shown.
The angles $\varphi$ and $\theta$ are used to rotate the simulated galaxy as if it was observed from the peculiar point of view yielding the imposed clustercentric distance $r_p$ and the recessional velocity $v_p$.

\bsp	%
\label{lastpage}
\end{document}